\documentclass[reprint, aps, prl, superscriptaddress, email]{revtex4-2}

\usepackage[utf8]{inputenc}
\usepackage[english]{babel}

\usepackage[T1]{fontenc}
\usepackage{dsfont}
\usepackage{siunitx}

\usepackage{mathtools}
\usepackage{amssymb}
\usepackage{braket}

\usepackage{graphicx}
\usepackage{xcolor}

\usepackage{array}
\usepackage{multirow}
\usepackage{dcolumn}

\usepackage{enumitem}

\usepackage[hidelinks]{hyperref}
    \hypersetup{
        colorlinks=true,
        linkcolor=blue,
        filecolor=blue,      
        urlcolor=blue,
        citecolor=blue
    }

\definecolor{white}{rgb}{1.0, 1.0, 1.0}
\definecolor{black}{rgb}{0.0, 0.0, 0.0}
\definecolor{red}{rgb}{0.8, 0.2, 0.2}
\definecolor{blue}{rgb}{0.0, 0.3, 0.7}
\definecolor{green}{rgb}{0.2, 0.7, 0.2}
\definecolor{yellow}{rgb}{1.0, 0.9, 0.2}
\definecolor{purple}{rgb}{0.6, 0.0, 0.6}
\definecolor{orange}{rgb}{1.0, 0.6, 0.0}

\renewcommand{\d}{\mathrm{d}} % infinitesimal
\newcommand{\e}{\mathrm{e}} % exponential
\newcommand{\qo}[1]{#1} % operator
\newcommand{\ex}[1]{\smash{\langle#1\rangle}} % expected value

\newcommand{\op}[2]{\smash{\ket{#1}\!\!\bra{#2}}} % outer product
\renewcommand{\H}{\qo{H}} % Hamiltonian
\begin{document}

\preprint{APS/123-QED}

% -------------------------------------------
% [-] Title
% -------------------------------------------

\title{Spectral Signatures of Vibronic Coupling in Trapped Cold Ionic Rydberg Systems}

\author{Joseph W. P. Wilkinson}
\affiliation{Institut f{\"u}r Theoretische Physik, Universit{\"a}t T{\"u}bingen, Auf der Morgenstelle 14, 72076 T{\"u}bingen, Germany}

\author{Weibin Li}
\affiliation{School of Physics and Astronomy and Centre for the Mathematics and Theoretical Physics of Quantum Non-Equilibrium Systems, The University of Nottingham, Nottingham, NG7 2RD, United Kingdom}

\author{Igor Lesanovsky}
\affiliation{Institut f{\"u}r Theoretische Physik, Universit{\"a}t T{\"u}bingen, Auf der Morgenstelle 14, 72076 T{\"u}bingen, Germany}
\affiliation{School of Physics and Astronomy and Centre for the Mathematics and Theoretical Physics of Quantum Non-Equilibrium Systems, The University of Nottingham, Nottingham, NG7 2RD, United Kingdom}

\date{May 29, 2024}

% -------------------------------------------
% [-] Abstract
% -------------------------------------------

\begin{abstract} 
    Atoms and ions confined with electric and optical fields form the basis of many current quantum simulation and computing platforms.
    When excited to high-lying Rydberg states, long-ranged dipole interactions emerge which strongly couple the electronic and vibrational degrees of freedom through state-dependent forces.
    This vibronic coupling and the ensuing hybridization of internal and external degrees of freedom manifest through clear signatures in the many-body spectrum.
    We illustrate this by considering the case of two trapped Rydberg ions, for which the interaction between the relative vibrations and Rydberg states realizes a quantum Rabi model.
    We proceed to demonstrate that the aforementioned hybridization can be probed by radio frequency spectroscopy and discuss observable spectral signatures at finite temperatures and for larger ion crystals.
\end{abstract}

\maketitle

% -------------------------------------------
% [-] Introduction
% -------------------------------------------

\textit{Introduction.}---Systems of trapped ions have led to a number of breakthroughs in the fields of quantum many-body and non-equilibrium physics~\cite{lewenstein2007, gross2017, monroe2021}.
They have been used to study quantum phases of interacting spins~\cite{porras2004, friedenauer2008, schauss2015}, quantum phase transitions in open quantum many-body systems~\cite{greiner2002, vojta2003, diehl2008}, quantum thermodynamics principles~\cite{abah2012}, and molecular physics using Rydberg aggregates~\cite{wuster2011, schonleber2015, wuster2017}.
In conventional trapped ion quantum simulators, ions in energetically low-lying electronic states are employed to encode fictitious spin degrees of freedom (qubits)~\cite{kim2009, kim2010, barreiro2011, muller2011, schneider2012}.
Interactions and high-fidelity conditional operations are then mediated using a so-called phonon bus~\cite{bruzewicz2019, behrle2023}, the required spin-phonon or vibronic coupling being achieved by state-dependent light shifts~\cite{haljan2005}.
In a relatively recent development (see, e.g., Refs.~\cite{schmidt-kaler2011, feldker2015, labuhn2016, higgins2017, mokhberi2020, muller2008}), trapped ions have been excited to energetically high-lying electronic states, known as Rydberg states, that interact via electric dipole forces.
This mechanism allows for the implementation of strong coherent interactions, which have been utilized to generate submicrosecond entangling gate operations~\cite{zhang2020}, and to mediate effective spin interactions that do not rely on the phonon bus.
It also frees up the phonon degrees~of freedom, augmenting the trapped ion quantum simulator, facilitating the study of a range of interesting many-body phenomena in which trap vibrational modes are coupled to interacting electronic states~\cite{zhang2022, bharti2023, gambetta2020, gambetta2021, magoni2023}.

\begin{figure}[ht!]
    \centering
    \includegraphics[width=\columnwidth]{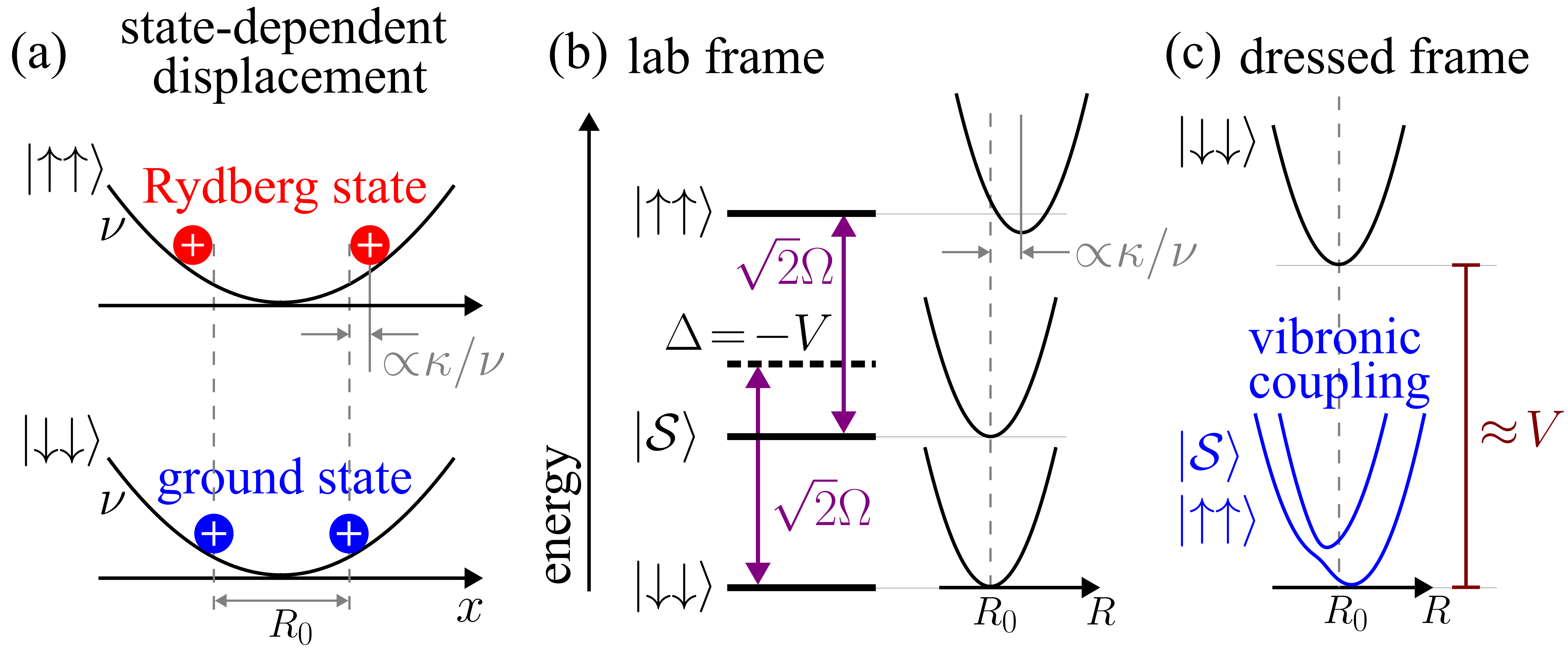}
    \caption{\textbf{System.}
        (a) Two ions confined within a harmonic potential with trap frequency $\nu$.
        When both ions are in their electronic ground states, i.e., $\ket{\downarrow\downarrow} = \ket{\downarrow} \otimes \ket{\downarrow}$, the equilibrium distance between the ions is $R_{0}$.
        However, when both ions are simultaneously excited to Rydberg states $\ket{\uparrow\uparrow}$, electric dipole interactions displace the ions from their equilibrium positions by an amount proportional to $\kappa / \nu$ where $\kappa$ parameterizes the strength of vibronic (i.e., spin-phonon) coupling.
        (b) Relevant energy levels for the system of trapped ions in the (stationary) lab frame.
        The laser, with detuning $\Delta$ and Rabi frequency $\Omega$, couples the state $\ket{\downarrow\downarrow}$, via the singly-excited symmetric state $\ket{\mathcal{S}} = [\ket{\uparrow\downarrow} + \ket{\downarrow\uparrow}] / \sqrt{2}$, to the doubly-excited (Rydberg) state $\ket{\uparrow\uparrow}$.
        We consider the regime where the laser detuning cancels the interaction between the Rydberg ions at their equilibrium separation $R_{0}$ (i.e., $\Delta = -V$).
        Electric dipolar forces between the Rydberg ions couple the electronic and relative vibrational motion.
        (c) External dynamics in the (rotating) dressed frame of the laser.
        In the state $\ket{\downarrow\downarrow}$, the ions experience a virtually unperturbed confinement, however, in the states $\ket{\mathcal{S}}$ and $\ket{\uparrow\uparrow}$, they hybridize with the relative motional degrees of freedom.
        The resulting coupled electronic potential surfaces are located at an energy of approximately $E \approx - V$.
    }
    \label{fig:model}
\end{figure}

In this work, we investigate a scenario where we create strong vibronic coupling in the electronic Rydberg state manifold between a pair of trapped ions.
This is achieved by exciting Rydberg states under so-called facilitation~or anti-blockade conditions~\cite{magoni2021, magoni2022, amthor2010, simonelli2016, letscher2017, helmrich2020, kitson2023, brady2023}.
Within this regime, the vibronic coupling between excited electronic states and phonons modes is described by a variant of the quantum Rabi model~\cite{xie2017}.
We show how the hybridized states can be experimentally probed via radio frequency modulation of the Rydberg state excitation laser, discuss the spectral signatures of the vibronic coupling, and also study their dependence on the temperature and number of ions.
Our investigation highlights the potential in using systems of trapped ions, or even atoms, excited to Rydberg states to realize complex scenarios with coupled electronic and vibrational motion that are of the utmost importance in, e.g., biological processes~\cite{rinaldi2014}, chemical reactions~\cite{hempel2018, schlawin2021, meng2023}, and molecular dynamics~\cite{macdonell2023, whitlow2023, valahu2023}. 

% -------------------------------------------
% [-] Model
% -------------------------------------------

\textit{Model.}---We consider a chain of ions trapped within a linear Paul trap. The internal degrees of freedom of each ion are modelled by two levels, denoted $\ket{\downarrow}$ and $\ket{\uparrow}$, that, respectively, represent an electronically low-lying ground state and high-lying excited Rydberg state of an alkaline earth metal ion~\cite{djerad1991}.
These states are coupled by a laser with Rabi frequency~$\Omega$ and detuning $\Delta$. The state $\ket{\uparrow}$~is assumed to be a dressed Rydberg state that is generated by coupling two suitably chosen states from the Rydberg manifold via a microwave (MW) field (see Refs.~\cite{mokhberi2020, muller2008, zhang2020}).
This dressing technique produces strong and controllable electric dipole-dipole interactions amongst Rydberg ions with a strength parameterized by $V \propto d^{2} / R_{0}^{3}$ with $d$ the electric transition dipole moment between the microwave coupled Rydberg states and $R_{0}$ the equilibrium distance between the ions~\footnote{See Supplemental Material at [URL] for a detailed derivation of the spin-phonon coupled model Hamiltonian and values for experimental and theoretical parameters, which additionally includes Refs.~\cite{gallagher1988, gallagher1994, gallagher2023, paul1990, wineland1998, leibfried2003, major2005, hucul2008, foot2004, aymar1996, berkland1998, cook1985, griffiths2017, griffiths2018, friedrich2017, saffman2010, bollinger1994, enzer2000, fishman2008, kleczewski2012, williams2013, inlek2017, kolbe2012, bachor2016, weber2017, louck2023, li2014, li2013}.}.
The interaction amongst Rydberg states also gives rise to mechanical forces that, as shown in Fig.~\ref{fig:model}a, induce state-dependent displacements~\cite{li2012, gambetta2020}.
Note that mechanical effects are also present when single trapped ions are excited into Rydberg states~\cite{higgins2017, vogel2019}.~For
simplicity, we will not account for these here as they can be eliminated through precise control of the polarizability of the MW dressed Rydberg states~\cite{pokorny2020}.

To illustrate our ideas, we initially consider a system that consists of two ions, as depicted in Fig.~\ref{fig:model}a, and later generalize to many ions.
For brevity, we only outline the derivation of the spin-phonon coupled Hamiltonian here, and reserve further relevant details to the Supplemental Material (SM)~\cite{Note1}.
The model Hamiltonian for a system of trapped Rydberg ions is given by (n.b., $\hbar = 1$),
\begin{equation}
    \H = \sum_{i = 1}^{2} \qo{h}_{i} + V_{12} \qo{n}_{1} \qo{n}_{2} + \omega_{2} \qo{a}_{2}^{\dagger} \qo{a}_{2}, \quad
    \qo{h}_{i} = \Delta \qo{n}_{i} + \Omega \qo{\sigma}_{i}^{x},
\end{equation}
where $\qo{n}_{i} = \op{\uparrow}{\uparrow}_{i}$ is the projector onto the Rydberg state of ion $i$ and $\qo{\sigma}_{i}^{x} = \op{\uparrow}{\downarrow}_{i} + \op{\downarrow}{\uparrow}_{i}$ the associated spin-flip operator.
The first two terms describe the effective spin dynamics modelling the ions' internal electronic degrees of freedom, the former the interactions of the ions with the electric field, and the latter the interactions between the ions in the Rydberg states via the distance-dependent potential $V_{12} = V(R_{12})$ with $R_{12}$ the interionic distance.
The final term governs the external vibrational degrees~of freedom, which are modelled by a single phonon mode of frequency $\omega_2$ with creation and annihilation operators $\qo{a}^{\dagger}_{2}$ and~$\qo{a}_{2}$.
In order to obtain a leading order coupling term, we linearly expand the dipole-dipole interaction potential $V(R_{12})$ about the equilibrium separation $R_{0}$ between the ions~\cite{mazza2020, magoni2021, magoni2022, magoni2023}.
Expressing the displacements of the ions about their equilibrium positions in terms of the phonon mode operators we get $V_{12} \approx V + \sum_{\smash{p = 1}}^{\smash{2}} \kappa_{p} [\qo{a}_{p}^{\dagger} + \qo{a}_{p}]$~with the spin-phonon coupling strength given by,
\begin{equation}\label{eq:coupling-strength}
    \kappa_{p} = - \frac{3 V}{R_{0}} \frac{\Gamma_{p}}{\sqrt{2 M \omega_{\smash{p}}}}, \quad
    V = \frac{1}{4 \pi \epsilon_{0}} \frac{d^{2}}{R_{0}^{3}}.
\end{equation}
Here, $M$ is the ion mass and $\Gamma_{p}$ the coupling coefficient associated to the phonon mode $p$ with frequency $\omega_{p}$.~For
two ions, state-dependent forces only couple the relative vibrational motion with the electronic dynamics.~In~terms
of the ion trap frequency $\nu$ (see Fig.~\ref{fig:model}a), the frequency of the relative mode is $\omega_{2} = \sqrt{3} \nu$ and the coupling strength $\kappa_{2} < 0$ since $\Gamma_{2} = \sqrt{2}$.
In contrast, for the center of mass mode we have $\omega_{1} = \nu$, yet $\kappa_{1} = 0$ as $\Gamma_{1} = 0$.
Accordingly, the Hamiltonian reads (see Eq.~(S61) in the SM~\cite{Note1}),
\begin{equation}\label{eq:two-ion-hamiltonian}
    \H = \sum_{i = 1}^{2} \qo{h}_{i} + V \qo{n}_{1} \qo{n}_{2} + \omega_{2} \qo{a}_{2}^{\dagger} \qo{a}_{2} + \kappa_{2} [\qo{a}_{2}^{\dagger} + \qo{a}_{2}] \qo{n}_{1} \qo{n}_{2}.
\end{equation}

\begin{figure}[t]
    \centering
    \includegraphics[width=\columnwidth]{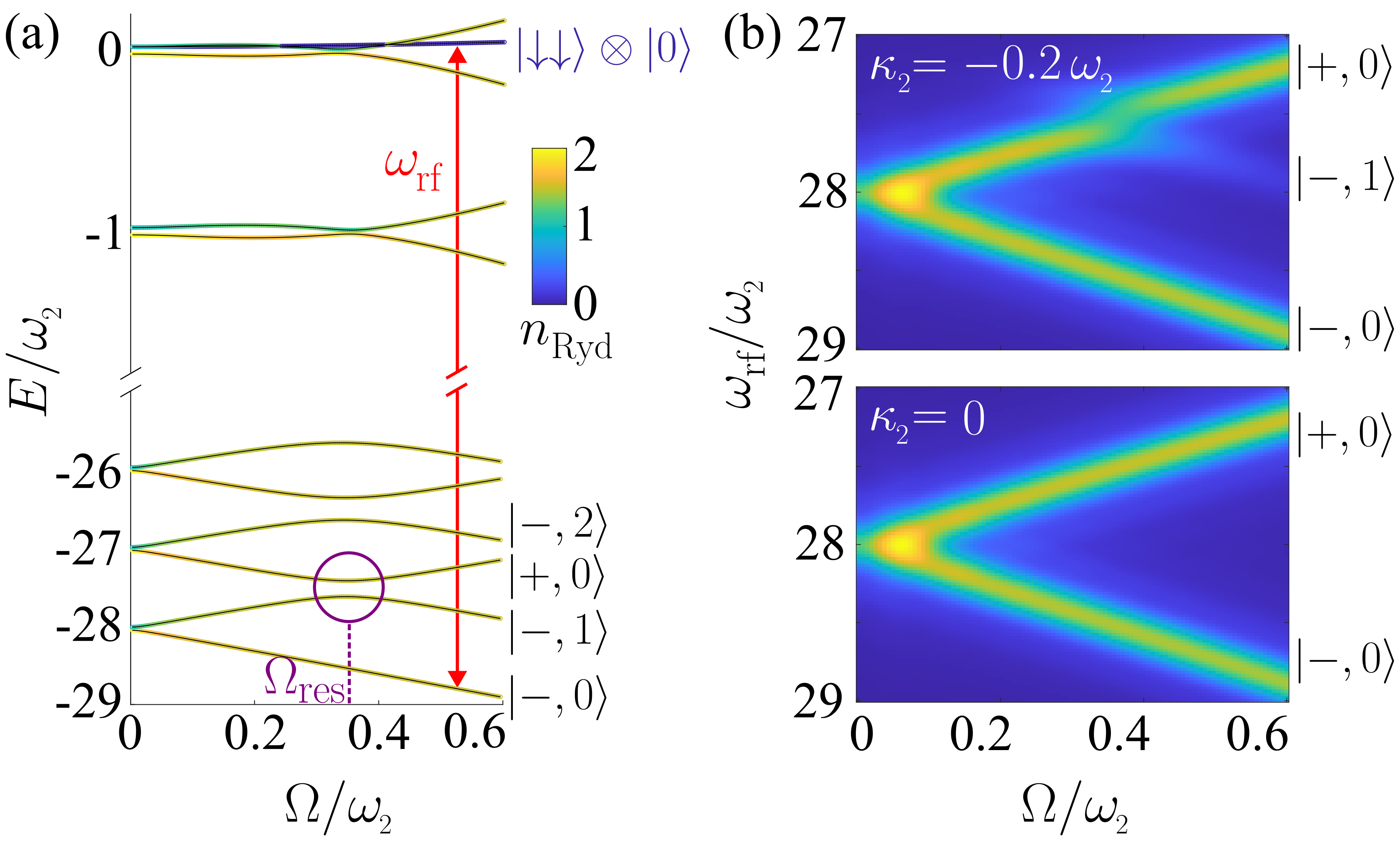}
    \caption{\textbf{Spectrum and radio frequency spectroscopy.}
        (a) Energy spectrum of the two-ion Hamiltonian in Eq.~\eqref{eq:two-ion-hamiltonian}~for $V = -\Delta = 28 \omega_2$ with $\kappa_{2} = -0.20 \omega_{2}$ as a function of the Rabi frequency $\Omega$.
        The color of the line encodes the average number of Rydberg excitations $n_{\mathrm{Ryd}} = \ex{\qo{n}_{1} + \qo{n}_{2}}$ of the approximate eigenstate.
        The blue line represents the initial state $\ket{\downarrow\downarrow} \otimes \ket{0}$ which is adiabatically connected to the electronic state $\ket{\downarrow\downarrow}$~in the limit $\kappa_{2} \to 0$ since, for all $\Omega$ considered, it contains only~a tiny admixture of the Rydberg states $\ket{\uparrow}$.
        Transitions between states are driven by applying a radio frequency (rf) field with frequency $\omega_{\mathrm{rf}}$.
        This facilitates the probing of the coupling that occurs in the vicinity of the resonance at $\Omega = \Omega_{\mathrm{res}} = \omega_{2} / 2 \sqrt{2}$, marked by the purple circle.
        Note that the states in the lower branches denote eigenstates in the limit $\kappa_{2} \to 0$ (see the main text for details).
        (b) Spectroscopy of the hybridized electronic and vibrational states.
        The system is initially prepared in the state $\ket{\downarrow\downarrow} \otimes \ket{0}$ for fixed $\Omega$.
        Irradiating the ions with an rf~field of frequency $\omega_{\mathrm{rf}}$ with strength $\Omega_{\mathrm{rf}} = 0.1 \omega_{2}$ (cf. Eq.~\eqref{eq:rabi-frequency}) and integrating the average number of Rydberg excitations over a period $\omega_{2} \tau = 30$ yields the signal shown.
        In the upper panel, where $\kappa_{2} = -0.20 \omega_{2}$, the hybridization clearly manifests as~an avoided crossing.
        This is in contrast to the lower panel, where $\kappa_{2} = 0$, and the electronic and vibrational motion decouple.
    }
    \label{fig:spectrum}
\end{figure}

The strength of the spin-phonon coupling $\kappa_{p}$ scales as $\kappa_{p} \sim M^{5/6}\nu^{13/6}$, therefore, the heavier the ion and larger the trap frequency, the stronger the coupling between the electronic and vibrational motion~\cite{Note1}.
For two ions, this is why we consider barium $\smash{^{138}\mathrm{Ba}^{+}}$ ions of isotopic mass $M = 137.9 \, \mathrm{u}$ as opposed to strontium $\smash{^{88}\mathrm{Sr}^{+}}$ ($M = 87.9 \, \mathrm{u}$) or calcium $\smash{^{40}\mathrm{Ca}^{+}}$ ($M = 40.0 \, \mathrm{u}$) ions which are currently used in trapped Rydberg ion experiments~\cite{mokhberi2020}.
Here, the electronically low-lying ground state $\ket{\downarrow}$ is the metastable state $\ket{5\mathrm{D}_{\smash{5/2}}}$, whilst the highly-excited dressed Rydberg state $\ket{\uparrow}$ is a superposition $\ket{\uparrow} = [\ket{n\mathrm{P}_{\smash{1/2}}} - \ket{n\mathrm{S}_{\smash{1/2}}}] / \sqrt{2}$.
These two states are coupled by a two-photon excitation scheme via the intermediate state $\ket{7\mathrm{P}_{3/2}}$ \cite{mokhberi2020, zhang2020}.
Using Rydberg states with a principal quantum number $n = 60$ and linear Paul trap with frequency $\nu = 2 \pi \times 6 \, \mathrm{MHz}$, we obtain an equilibrium ion separation $R_{0} = 1.12 \, \unit{\micro\meter}$ which returns an interaction strength $V = 28 \omega_{2}$ and a coupling strength $\kappa_{2} = -0.20 \omega_{2}$ (see Fig.~\ref{fig:spectrum}b).
We note that these values are somewhat extreme, yet feasible \cite{mokhberi2020}.
Later, we will show that these can be relaxed significantly to more typical values when considering larger ion crystals.

% -------------------------------------------
% [-] Spectrum
% -------------------------------------------

\textit{Spectrum.}---In the following, we consider the situation in which the dynamics is subject to the facilitation (anti-blockade) constraint where the laser detuning $\Delta$ cancels the interaction energy $V$ (i.e., $\Delta + V = 0$), as illustrated in Fig.~\ref{fig:model}b.
In this regime, the spin-phonon coupling is particularly prominent and a simplified analytical model can be developed.
Due to the level symmetry, the laser only couples the unexcited state $\ket{\downarrow\downarrow}$, the singly-excited symmetric state $\ket{\mathcal{S}} = [\ket{\uparrow\downarrow} + \ket{\downarrow\uparrow}] / \sqrt{2}$, and the doubly-excited Rydberg state $\ket{\uparrow\uparrow}$ (see Fig.~\ref{fig:model}b), with the singly-excited antisymmetric state $\ket{\mathcal{A}} = [\ket{\downarrow\uparrow} - \ket{\uparrow\downarrow}] / \sqrt{2}$ decoupled from the aforementioned dynamics.
Taking into account that the interaction energy $V \gg \Omega$, we note that the state $\ket{\downarrow\downarrow}$ only acquires a weak light shift and so can similarly be neglected.
On the other hand, the states $\ket{\mathcal{S}}$ and $\ket{\uparrow\uparrow}$ are resonantly coupled to the laser field with the electronic state $\ket{\uparrow\uparrow}$ also coupled to the vibrational~mode.
The approximate Hamiltonian is then (see SM~\cite{Note1}),
\begin{equation}
    \H =
    \begin{bmatrix}
        -V & \sqrt{2} \Omega \\
        \sqrt{2} \Omega & -V \\
    \end{bmatrix}
    + \omega_{2} \qo{a}_{2}^{\dagger} \qo{a}_{2} + \kappa_{2} \!
    \begin{bmatrix}
        1  & 0 \\
        0 & 0 \\
    \end{bmatrix}
    \! [\qo{a}_{2}^{\dagger} + \qo{a}_{2}],
\end{equation}
where the energy of the hybridized states is with respect to the state $\ket{\downarrow\downarrow}$, as pictured in Fig.~\ref{fig:model}c.

In Fig.~\ref{fig:spectrum}a, we show the full vibronic coupled spectrum for $V = 28 \omega_{2}$ and $\kappa_{2} = -0.20 \omega_{2}$ as a function of the laser Rabi frequency $\Omega$.
In the region with energy $E \approx -V$,~we indeed observe an avoided crossing, indicated by a circle, which is a manifestation of the strong coupling between the internal electronic and external vibrational degrees~of freedom.
In order to study this coupling, we remark that the interaction strength $V \gg \kappa_{2}$ which allows us to treat the spin-phonon coupling as a perturbation.
Introducing the following electronic eigenstates $\ket{\pm} = [\ket{\uparrow\uparrow} \pm \ket{\mathcal{S}}] / \sqrt{2}$ of the unperturbed Hamiltonian (i.e., for $\kappa_{2} = 0$), we can then rewrite the approximate model Hamiltonian as, 
\begin{equation}\label{eq:resonant-hamiltonian}
\begin{aligned}
    \H & =
    \begin{bmatrix}
        E_{+} & 0 \\
        0 & E_{-} \\
    \end{bmatrix}
    + \omega_{2} \qo{a}_{2}^{\dagger} \qo{a}_{2} + \frac{\kappa_{2}}{2} \!
    \begin{bmatrix}
        1 & 1 \\
        1 & 1 \\
    \end{bmatrix}
    \! [\qo{a}_{2}^{\dagger} + \qo{a}_{2}],
\end{aligned}
\end{equation}
with $E_{\pm} = -V \pm \sqrt{2} \Omega$ the electronic energy eigenvalues.
This Hamiltonian is a variant of the quantum Rabi model with spin-phonon coupling constant $\kappa_{2}$~\cite{xie2017}.
For $\kappa_{2} = 0$, the spin-phonon dynamics decouple and the Hamiltonian becomes diagonal.
The corresponding energy eigenvalues are $E_{\pm, N} = E_{\pm} + N \omega_{2}$, whilst the associated eigenstates are $\ket{\pm, N} = \ket{\pm} \otimes \ket{N}$, where $\ket{N}$ is an eigenstate of the number operator with eigenvalue~$N$.
A resonance occurs when any pair of these energies becomes degenerate, e.g., the resonance shown in Fig.~\ref{fig:spectrum}a is due to states $\ket{+, 0}$ and $\ket{-, 1}$, which become degenerate at $\Omega = \Omega_{\mathrm{res}} \approx \omega_{2} / 2 \sqrt{2}$.
Notice that this is only an estimate for the value of~the resonance frequency $\Omega_{\mathrm{res}}$, since we are neglecting second order light shifts.
In general, resonances occur whenever the Rabi frequency $\Omega = \Omega_{\mathrm{res}} \approx N \omega_{2} / 2 \sqrt{2}$ with $N \in \mathbb{N}$.~If
we calculate the approximate eigenstates at the resonance between the states $\ket{+, 0}$ and $\ket{-, 1}$ highlighted in Fig.~\ref{fig:spectrum}a, we find that~\cite{Note1},
\begin{equation}\label{eq:hybrid-state}
    \ket{E_{\pm}^{\mathrm{res}}} = \frac{1}{2} \big[\!\ket{\uparrow\uparrow} \otimes [\ket{1} \pm \ket{0}] - \ket{\mathcal{S}} \otimes [\ket{1} \mp \ket{0}]\big],
\end{equation}
which evidently shows hybridization of the electronic and vibrational degrees of freedom.
The resonant energy level splitting is given by the coupling strength $\kappa_{2}$.

% -------------------------------------------
% [-] Spectroscopy
% -------------------------------------------

\begin{figure}[t]
    \centering
    \includegraphics[width=\columnwidth]{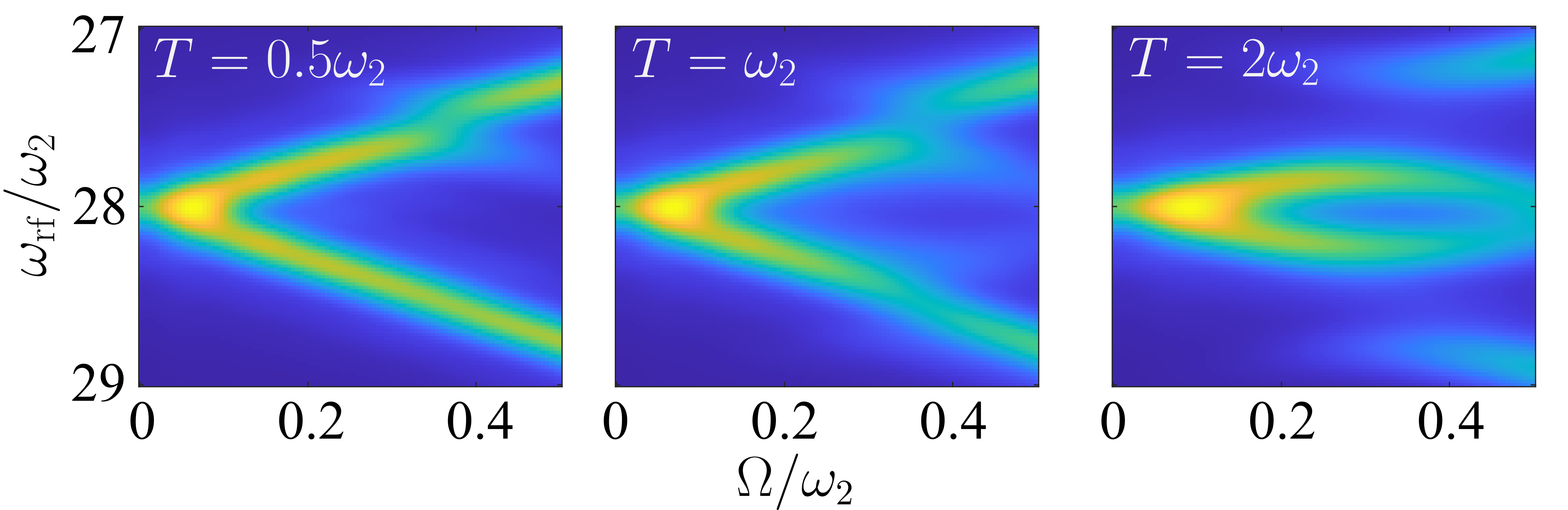}
    \caption{\textbf{Spectroscopy at finite temperature.}
        Radio~frequency spectra for the initially prepared state $\ket{\downarrow\downarrow}\!\bra{\downarrow\downarrow} \otimes \rho_T$~at different temperatures $T$.
        Initial states with high vibrational mode quantum numbers $n$ couple to more symmetric parts~of the spectrum.
        As such, the higher the temperature, the more symmetric the signal becomes about $\omega_{\mathrm{rf}} = V$.
        The data plotted is generated using the parameters given in Fig.~\ref{fig:spectrum} with the coupling strength $\kappa_{2} = - 0.20 \omega_{2}$ (see Fig.~\ref{fig:spectrum}b for details).~Note
        that for $\kappa_{2} = 0$, all these plots would be indistinguishable~from the bottom panel of Fig.~\ref{fig:spectrum}b.
    }
    \label{fig:spectrum-finite-temperature}
\end{figure}

\textit{Spectroscopy}.---In order to probe the energy spectrum shown in Fig.~\ref{fig:spectrum}a in an experiment, we propose to perform radio frequency (rf) spectroscopy.
To implement this,~we replace the Rabi frequency in Eq.~\eqref{eq:two-ion-hamiltonian} according to,
\begin{equation}\label{eq:rabi-frequency}
    \Omega \to \Omega(t) = \Omega + \Omega_{\mathrm{rf}} \cos(\omega_{\mathrm{rf}} t),
\end{equation}
where $\omega_{\mathrm{rf}}$ and $\Omega_{\mathrm{rf}}$ are the radio frequency and amplitude modulation of the field.
The spectroscopic protocol is as follows.
To start, we prepare the system in the unexcited state $\ket{\downarrow\downarrow} \otimes \ket{0}$, i.e., the state within which both the spins and the phonon are, respectively, in their electronic and vibrational ground states.
Next, we switch on the laser~to set the desired value for the time-independent part of the Rabi frequency (i.e., $\Omega \neq 0$ and $\Omega_{\mathrm{rf}} = 0$).
Assuming that this proceeds adiabatically, this amounts to moving along the blue line in Fig.~\ref{fig:spectrum}a.
Note, however, that in practice, a sudden turning on of the laser should also suffice, since for all considered values of the Rabi frequency the state colored in blue corresponds to the initial state $\ket{\downarrow\downarrow} \otimes \ket{0}$, up to corrections of order $[\Omega / V]^{2}$.
Now the rf modulation is switched on (i.e., $\Omega_{\mathrm{rf}} \neq 0$) and, if the radio frequency $\omega_{\mathrm{rf}}$ is set to the energy splitting between two hybridized levels, illustrated by the red arrow in Fig.~\ref{fig:spectrum}a, a transition occurs.
Given that the initial state contains no Rydberg excitations, monitoring the number of ions that are in Rydberg states provides a direct spectroscopic signature of whether a transition has taken place, as demonstrated in Fig.~\ref{fig:spectrum}b, where we plot the time-integrated number of Rydberg excitations $I = \int_{0}^{\tau} \d t \, \ex{\qo{n}_{1} + \qo{n}_{2}}(t)$ as a function of $\omega_{\mathrm{rf}}$ and $\Omega(t)$ over the time interval $\omega_{2} \tau = 30$.

Transitions can only occur if the Hamiltonian possesses a non-vanishing matrix element between initial and final states.
For the chosen initial state $\ket{\downarrow\downarrow} \otimes \ket{0}$, this is only the case if the final state contains some admixture of the state $\ket{\mathcal{S}} \otimes \ket{0}$.
Hence, in the limit $\kappa_{2} = 0$, only the states $\ket{\pm, 0}$ can be excited, as demonstrated in the lower panel of Fig.~\ref{fig:spectrum}b.
However, with increased vibronic coupling the electronic and vibrational motion hybridize such that, in the vicinity of the resonance denoted in Fig.~\ref{fig:spectrum}a, the state is approximated by that in Eq.~\eqref{eq:hybrid-state}.
Given that this state exhibits overlap with the state $\ket{\mathcal{S}} \otimes \ket{0}$, it can be excited from the initial state and, from inspection of Fig.~\ref{fig:spectrum}b, one clearly observes the associated avoided crossing.

At finite temperature, the initial phonon state is a thermal state, $\rho_{T} = \sum_{N = 0}^{\infty} \e^{-N \omega_{2} / T} \!/ [1 - \e^{- \omega_{2} / T}] \op{N}{N}$.
The occupation of these higher vibrational states opens novel transition channels.
Indeed, in contrast to the case where $T = 0$, these aforementioned transitions do not probe the lower edge of the spectrum, delimited by the state $\ket{-, 0}$ (see Fig.~\ref{fig:spectrum}b), whose energy decreases linearly with $\Omega$.
Rather, they lead to states being symmetrically repelled by other states of higher and lower energy.
For example, the initial state in Fig.~\ref{fig:spectrum}a couples to states with asymptotes $\ket{+, 0}$ and $\ket{-, 2}$.
This coupling to more symmetric parts of the spectrum manifests in a spectroscopic signal, as pictured in Fig.~\ref{fig:spectrum-finite-temperature}.
For sufficiently low $T$, the signal is similar to that in the upper panel in Fig.~\ref{fig:spectrum}b.
However, as the temperature increases the signal becomes symmetric about $\omega_{\mathrm{rf}} = V$.
Note that without spin-phonon coupling, the spectrum would be identical to that in the lower panel of Fig.~\ref{fig:spectrum}b for all $T$.
Hence, small, but finite temperatures increase the spectral signature of the vibronic coupling.

% -------------------------------------------
% [-] Ion crystals 
% -------------------------------------------

\begin{figure}[t]
    \centering
    \includegraphics[width=0.9\columnwidth]{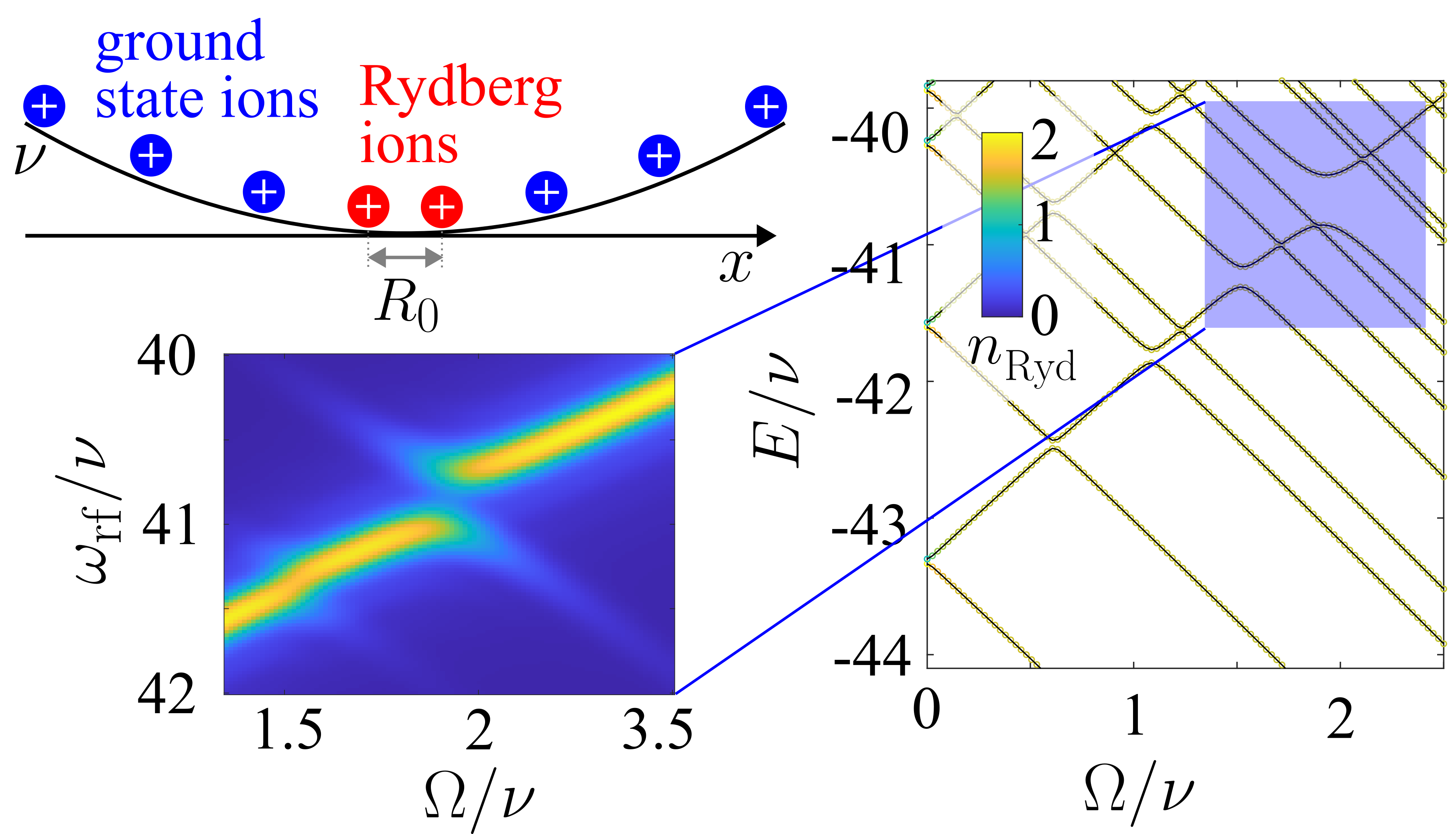}
    \caption{\textbf{Larger ion crystals.}
        Energy and radio frequency spectrum of a chain of trapped Rydberg ions.
        As the number of ions in the crystal $N$ increases, the equilibrium separation between the two centermost ions $R_{0}$ decreases~\cite{james1998}.
        Here, we consider a chain of $N = 8$ strontium $\smash{^{88}\mathrm{Sr}^{+}}$ ions confined in a trap with frequency $\nu = 2 \pi \times 2 \, \mathrm{MHz}$.
        In the electronic~ground state $\ket{\downarrow \cdots \downarrow}$, the equilibrium separation between the central ions $R_{0} = 1.37 \, \unit{\micro\meter}$.
        When the centermost ions are excited~to the Rydberg states $\ket{\uparrow}$, they interact~with strength $V = 43 \nu$.
        In contrast to the case of two ions (cf. Fig.~\ref{fig:spectrum}), the two spins couple to four phonons of frequency $\omega_{2} = 1.73 \nu$, $\omega_{4} = 3.06 \nu$, $\omega_{6} = 4.29 \nu$, $\omega_{8} = 5.44 \nu$ with coupling strength $\kappa_{2} = -0.06 \nu$, $\kappa_{4} = -0.10 \nu$, $\kappa_{6} = -0.15 \nu$, $\kappa_{8} = -0.27 \nu$.
        Note, in particular, the coupling to the latter mode which manifests as an avoided crossing that can be probed via radio frequency spectroscopy (cf. Fig.~\ref{fig:spectrum}b), as demonstrated in the outset.
    }
    \label{fig:ion-crystal}
\end{figure}

\textit{Ion crystals}.---We now generalize our considerations to a chain of $N$ ions confined within a linear Paul trap~\cite{muller2008}.
For simplicity, we assume that only the centermost pair of ions are irradiated with the laser such that the internal electronic degrees of freedom of the unexcited ions decouple from the many-body spin-phonon coupled dynamics.
This leads to the following Hamiltonian~\cite{Note1},
\begin{equation}\label{eq:many-ion-hamiltonian}
    \ \ \H = \sum_{i = 1}^{2} \qo{h}_{i} + V \qo{n}_{1} \qo{n}_{2} + \sum_{p = 1}^{N} [\omega_{p} \qo{a}_{p}^{\dagger} \qo{a}_{p} + \kappa_{p} [\qo{a}_{p}^{\dagger} + \qo{a}_{p}] \qo{n}_{1} \qo{n}_{2}], \ \
\end{equation}
with the former two terms corresponding to the electronic motion defined as in Eq.~\eqref{eq:two-ion-hamiltonian} and where for simplicity the centermost pair of ions are labelled by $i = 1, 2$.
The~latter terms then respectively describe the external and coupled motion, with $\omega_{p}$ the frequency of the phonon mode $p$ and $\kappa_{p}$ the associated strength of the coupling to the internal dynamics.
Note, for even numbers of ions $N$, the coupling coefficients $\Gamma_{p}$ and, consequently, the coupling strengths $\kappa_{p}$ [see Eq.~\eqref{eq:coupling-strength}] for modes with odd $p$ vanish.
Hence, the corresponding modes decouple and can be neglected. 

Larger ion crystals give rise to increased spin-phonon coupling strengths since ions in the trap center get closer and their interaction features stronger forces (see Fig. \ref{fig:ion-crystal}).
To demonstrate this we consider an ion crystal of~$N = 8$ strontium $^{\smash{88}}\mathrm{Sr}^{\smash{+}}$ ions of mass $M = 87.9 \, \mathrm{u}$ with parameter values that are significantly relaxed compared to the case of $N = 2$ barium $^{\smash{138}}\mathrm{Ba}^{\smash{+}}$ ions considered initially.
Here, the state $\ket{\downarrow} = \ket{4\mathrm{D}_{\smash{5/2}}}$ is a metastable state, whilst the state $\ket{\uparrow} = [\ket{n\mathrm{P}_{\smash{1/2}}} - \ket{n\mathrm{S}_{\smash{1/2}}}] / \sqrt{2}$ is a (dressed) Rydberg state with principal quantum number $n = 50$.
With trap frequency $\nu = 2 \pi \times 2 \, \mathrm{MHz}$, the equilibrium separation of the centermost ions $R_{0} = 1.37 \, \smash{\unit{\micro\meter}}$ and the corresponding interaction strength $V = 43 \nu$.
In contrast to the two ion case (see Fig.~\ref{fig:model}a), the spins now couple to four phonons, with frequencies and coupling strengths that are listed in Fig.~\ref{fig:ion-crystal}.
Here, the coupling to the $p = 8$ mode manifests~in Fig.~\ref{fig:ion-crystal} as a clearly observable avoided crossing.
Note that all parameters used are tabulated in the SM~\cite{Note1}.

% -------------------------------------------
% [-] Conclusions
% -------------------------------------------

\textit{Summary and outlook}.---In this work, we demonstrate that strong state-dependent forces in Rydberg ions allow for the engineering and exploring of vibronic interactions in trapped ion quantum simulators.
Spectral signatures of coupling between the electronic and vibrational motion are directly visible in the spectroscopy of Rydberg states with radio frequency modulated laser.
Whilst we focused on analytically and numerically tractable situations, the exponential growth of the number of degrees of freedom rapidly allows one to reach many-body scenarios that~are intractable on classical computers.
Spatially resolved and quantitative Rydberg state spectroscopy in the precisely controllable environment of such augmented trapped ion quantum simulation platforms can be used to benchmark and advance numerical approximations schemes, e.g., by facilitating an understanding of which quantum correlations are most important to capture the observed spectral signatures as the number of degrees of freedom grows.

% -------------------------------------------
% [-] Acknowledgements
% -------------------------------------------

\begin{acknowledgments}
    We gratefully acknowledge discussions with M. Hennrich.
    We are grateful for funding from the Deutsche Forschungsgemeinschaft (DFG, German Research Foundation) under Projects No. 428276754 and 435696605 as well as through the Research Unit FOR 5413/1, Grant No. 465199066.
    This project has also received funding from the European Union’s Horizon Europe research and innovation program under Grant Agreement No. 101046968 (BRISQ).
    This work was supported by the University of Nottingham and the University of T{\"u}bingen’s funding as part of the Excellence Strategy of the German Federal and State Governments, in close collaboration with the University of Nottingham.
    This work also received supported from the Engineering and Physical Sciences Research Council [grant numbers EP/V031201/1 and EP/W015641/1].
    JW was supported by the University of T{\"u}bingen through a Research@T{\"u}bingen fellowship.
    The code used to produce the data supporting the findings of this article is available on Zenodo~\cite{zenodo2024}.
\end{acknowledgments}

% -------------------------------------------
% [-] Bibliography
% -------------------------------------------

\bibliography{main.bib}

\end{document}

\preprint{APS/123-QED}

\widetext

\thispagestyle{empty}

% -------------------------------------------
% [-] Title
% -------------------------------------------

\begin{center}
    {\Large{
        \MakeUppercase{Supplemental Material}
    }\par}
    \vskip 18pt
    {\large{
        \textbf{Spectral Signatures of Vibronic Coupling in Trapped Cold Ionic Rydberg Systems}
    }\par}
    \vskip 12pt
    {\normalsize{
        Joseph W. P. Wilkinson\textsuperscript{1},
        Weibin Li\textsuperscript{2}, and
        Igor Lesanovsky\textsuperscript{1,2}
    }\par}
    \vskip 6pt
    {\small{
        \textsuperscript{1}\!\!\textit{
            Institut f{\"u}r Theoretische Physik,
            Universit{\"a}t T{\"u}bingen,
            Auf der Morgenstelle 14,
            72076 T{\"u}bingen,
            Germany
        } \\
        \textsuperscript{2}\!\!\textit{
            School of Physics and Astronomy and Centre for the Mathematics and Theoretical Physics of Quantum Non-Equilibrium Systems,
            University of Nottingham,
            Nottingham, NG7 2RD,
            United Kingdom
        }
    }\par}
    \vskip 0pt
    {\small{
        (Dated: May 29, 2024)
    }}
\end{center}

% -------------------------------------------
% [-] Introduction
% -------------------------------------------

In this supplemental material, we provide some useful information complementing the main text, including detailed calculations and derivations of important expressions and explicit values for relevant quantities and parameters.
\begin{itemize}
    \item In Sec.~\ref{sec:hamiltonian}, we derive the model Hamiltonian in Eq.~(8), and hence Eqs.~(1) and~(3), of the main text describing the spin-phonon coupled dynamics of the trapped Rydberg ion crystal from a more general many-body Hamiltonian.
    Additionally, we derive the variant of the quantum Rabi model Hamiltonian in Eq.~(5), and so Eq.~(4), of the main text, as well as the approximate eigenstates at the vibronic resonance in Eq.~(6).
    \item In Sec.~\ref{sec:phonons}, we tabulate values for the theoretical quantities and experimental parameters used in the main text. 
\end{itemize}

% -------------------------------------------
% [-] Body
% -------------------------------------------

\section{Derivation of the spin-phonon coupled model Hamiltonian}\label{sec:hamiltonian}

\subsection{Hamiltonian of a single trapped Rydberg ion}

In the main text, we consider a system of $N$ interacting alkali earth metal Rydberg ions confined within the electric quadrupolar potential of a linear Paul trap.
These ions possess a valence electron residing in an open outer shell that orbits in the modified Coulomb potential of a nucleus screened by the surrounding core electrons occupying the closed inner shells~\cite{djerad1991}.
In order to solve and subsequently model the dynamics of such a system, we must necessarily employ an approximation which reduces the intractable many-body problem to a solvable two-body problem~\cite{gallagher1988, gallagher1994, gallagher2023}.
For alkali earth metal ions, this is feasible since the inner electrons form closed shells around the nucleus.
As such, the ions can be modelled as consisting of an ionic core, composed of a nucleus screened by core electrons, with charge~$2e$, mass~$m_{\mathrm{c}}$, and position~$\r_{\mathrm{c}}$ and a valence electron with charge~$-e$, mass~$m_{\mathrm{e}}$, and position~$\r_{\mathrm{e}}$.

To confine the ions within a restricted region of space, a linear Paul trap~\cite{paul1990} is used which provides three-dimensional confinement for charged particles by applying static electric fields in combination with oscillating electric fields~\cite{wineland1998, leibfried2003, major2005}.
Together, these electric fields generate an approximately electric quadrupolar potential at the trap center of the form,
\begin{equation}
    \Phi(\qo{\r}, t) = \alpha \cos(\omega t) [\qo{r}_{x}^{2} - \qo{r}_{y}^{2}] + \beta [3 \qo{r}_{z}^{2} - \qo{\r}^{2}],
\end{equation}
where $\qo{\r} = (\qo{r}_{x}, \qo{r}_{y}, \qo{r}_{z})$ is the position operator of the trapped ion, $\omega$ is the radio frequency (RF) of the time-dependent driving field, whilst $\alpha$ and $\beta$ are the radial and axial electric field gradients, which are determined by the geometry of the trap electrodes and applied voltages~\cite{hucul2008}.
In order to facilitate transitions between the electronic energy levels, we additionally employ an electric potential associated to a time-dependent electric field $\bm{E}(\qo{\r}, t)$ which takes the form~of~a polychromatic plane wave.
Assuming only dipolar coupling of the applied electric field with the ionic core and valence electron, and within the dipole approximation~\cite{foot2004}, the Hamiltonian of the trapped Rydberg ion reads,
\begin{equation}\label{eq:rydberg-ion-hamiltonian}
    \H = \frac{\qo{\p}_{\mathrm{c}}^{2}}{2 m_{\mathrm{c}}} + \frac{\qo{\p}_{\mathrm{e}}^{2}}{2 m_{\mathrm{e}}} + V(\ab{\qo{\r}_{\mathrm{e}} - \qo{\r}_{\mathrm{c}}}) + 2 e \Phi(\qo{\r}_{\mathrm{c}}, t) - e \Phi(\qo{\r}_{\mathrm{e}}, t) + 2 e \qo{\r}_{\mathrm{c}} \cdot \bm{E}(t) - e \qo{\r}_{\mathrm{e}} \cdot \bm{E}(t),
\end{equation}
where $\qo{\r}_{\mathrm{c}}$, $\qo{\r}_{\mathrm{e}}$, $\qo{\p}_{\mathrm{c}}$, and $\qo{\p}_{\mathrm{e}}$ are the position and momentum operators of the ionic core and valence electron, respectively, and $V(\ab{\qo{\r}_{\mathrm{e}} - \qo{\r}_{\mathrm{c}}})$ the Coulomb potential approximating the interaction between the valence electron and ionic core~\cite{aymar1996}.
Introducing the center of mass and relative coordinates, defined by $\qo{\R} \approx \qo{\r}_{\mathrm{c}}$ and $\qo{\r} \approx \qo{\r}_{\mathrm{e}} - \qo{\r}_{\mathrm{c}}$, and transforming into a rotating frame oscillating at the RF driving frequency $\omega$ within the center of mass frame (see Ref.~\cite{muller2008} for a detailed derivation), the Hamiltonian in Eq.~\eqref{eq:rydberg-ion-hamiltonian} can be rewritten compactly as,
\begin{equation}
    \H = \H_{\mathrm{ex}} + \H_{\mathrm{in}} + \H_{\mathrm{ex\text{-}in}},
\end{equation}
where the Hamiltonian terms $\H_{\mathrm{ex}}$, $\H_{\mathrm{in}}$, and $\H_{\mathrm{ex\text{-}in}}$ respectively describe the external dynamics of the ionic core, the internal dynamics of the valence electron, and their coupled dynamics due to electric potential of the linear Paul trap.

The Hamiltonian governing the external dynamics of the trapped ionic core can be split into two terms,
\begin{equation}
    \H_{\mathrm{ex}} = \H_{\mathrm{ex}}^{\mathrm{sec}} + \H_{\mathrm{ex}}^{\mathrm{mic}},
\end{equation}
where $\H_{\mathrm{ex}}^{\mathrm{sec}}$ dictates the slow secular motion and $\H_{\mathrm{ex}}^{\mathrm{mic}}$ the fast driven motion called \textit{micromotion}~\cite{major2005}.
This separation, made manifest by the transformation into the oscillating frame, results from the composition of the static and rapidly oscillating fields that give rise to an effectively static \textit{ponderomotive} harmonic potential~\cite{berkland1998}. The former term reads,
\begin{equation}
    \H_{\mathrm{ex}}^{\mathrm{sec}} = \frac{1}{2 M} \sum_{u} [\qo{P}_{u}^{2} + M^{2} \nu_{u}^{2} \qo{R}_{u}^{2}],
\end{equation}
for $u = x, y, z$ where $\nu_{u}$ are the oscillation frequencies of the trapped ion within the effective harmonic potential,
\begin{equation}\label{eq:trap-frequencies}
    \nu_{x} = \nu_{y} = \sqrt{\frac{2 e^{2} \alpha^{2}}{M^{2} \omega^{2}} - \frac{2 e \beta}{M}},
    \qquad
    \nu_{z} = \sqrt{\frac{4 e \beta}{M}},
\end{equation}
while the latter term describing the micromotion is given by,
\begin{equation}
    \H_{\mathrm{ex}}^{\mathrm{mic}} = - \frac{2 e \alpha}{M \omega} \sin(\omega t) [\qo{R}_{x} \qo{P}_{x} - \qo{R}_{y} \qo{P}_{y}] - \frac{e^{2} \alpha^{2}}{M \omega^{2}} \cos(2 \omega t) [\qo{R}_{x}^{2} + \qo{R}_{y}^{2}] + e \qo{\R} \cdot \bm{E}(t).
\end{equation}
For typical experimental parameters (see, e.g., the review in Ref.~\cite{mokhberi2020}), the effects of the micromotion are negligible~as the micromotion, associated to fast driven motion, occurs on much shorter timescales than the slow secular motion~of the trapped ion~\cite{cook1985}.
Specifically, in order to neglect the micromotion we require $M \omega^{2} \gg e \alpha$ which is satisfied for~the experimental parameters considered.
Moreover, the micromotion term arising from the coupling with the applied field can similarly be neglected since it is far from being resonant with either the internal or external dynamics~\cite{muller2008}.

The Hamiltonian for the internal dynamics, which incorporates the motion of the orbiting valence electron within the modified Coulomb potential of the screened ionic core superposed with the electric quadrupolar potential of the linear Paul trap and electric dipolar potential of the applied field reads,
\begin{equation}
    \H_{\mathrm{in}} = \H_{\mathrm{in}}^{\mathrm{free}} + \H_{\mathrm{in}}^{\mathrm{trap}} + \H_{\mathrm{in}}^{\mathrm{field}}.
\end{equation}
Here, $\H_{\mathrm{in}}^{\mathrm{free}}$ contains the field-free electronic Hamiltonian of the valence electron and is given by,
\begin{equation}
    \H_{\mathrm{in}}^{\mathrm{free}} = \frac{\qo{\p}^{2}}{2 m} + V(\ab{\qo{\r}}),
\end{equation}
where $V(\ab{\qo{\r}})$ is a relativistic model potential dependent on the orbital angular momentum state of the valence electron approximating the modified hydrogenic Coulomb interaction experienced by the valence electron due to the screened nucleus and core electrons~\cite{schmidtkaler2011}.
The remaining terms, $\H_{\mathrm{in}}^{\mathrm{trap}}$ and $\H_{\mathrm{in}}^{\mathrm{field}}$, then describe the interaction of the valence electron with the electric potentials of the linear Paul trap and applied field,
\begin{equation}
    \H_{\mathrm{in}}^{\mathrm{trap}} = - e \alpha \cos(\omega t) [\qo{r}_{x}^{2} - \qo{r}_{y}^{2}] - e \beta [3 \qo{r}_{z}^{2} - \qo{\r}^{2}], \qquad
    \H_{\mathrm{in}}^{\mathrm{field}} = - e \qo{\r} \cdot \bm{E}(t),
\end{equation}
where corrections due to the finite mass of the ionic core have been neglected.
This approximation is well justified for the alkali earth metal ions considered here, since $m_{\mathrm{e}}/m_{\mathrm{c}} \approx 10^{-5}$ and so $M \approx m_{\mathrm{c}}$ and $m \approx m_{\mathrm{e}}$.

The electronic motion of the valence electron occurs on much faster timescales than the motion of the trapped ionic core, however, due to the electric quadrupolar potential of the linear Paul trap, the center of mass motion of the ionic core and relative motion of the valence electron are \textit{nonseparable}.
This intrinsic motion is accounted for accordingly by the Hamiltonian of the coupled dynamics,
\begin{equation}
    \H_{\mathrm{ex\text{-}in}} = \H_{\mathrm{ex\text{-}in}}^{\mathrm{trap}}, \qquad
    \H_{\mathrm{ex\text{-}in}}^{\mathrm{trap}} = - 2 e \alpha \cos(\omega t) [\qo{R}_{x} \qo{r}_{x} - \qo{R}_{y} \qo{r}_{y}] - 2 e \beta [3 \qo{R}_{z} \qo{r}_{z} - \qo{\R} \cdot \qo{\r}].
\end{equation}

\subsection{Interacting many-body Hamiltonian of a trapped Rydberg ion chain}

With an expression for the Hamiltonian of a trapped Rydberg ion, we now turn to the discussion of the interactions between ions.
As we will demonstrate, the relative positions of the ions (i.e., the distances between the ions) are small enough that we can neglect retardation effects due to the electric potentials~\cite{griffiths2017}, yet large enough that we can ignore exchange interactions between the valence electrons~\cite{griffiths2018}.
These assumptions drastically simplify the calculations and allow us to treat the valence electrons of the ions as \textit{distinguishable} particles.
Consequently, the Coulomb interactions between the charges of the ionic cores and valence electrons can be described using a multipole expansion~\cite{friedrich2017}.

The Coulomb interaction between the charges of the ionic cores and valences electrons of ions $i$ and $j$, with center of mass and relative positions $\R_{i}$ and $\R_{j}$ and $\r_{i}$ and $\r_{j}$, is given by,
\begin{equation}\label{eq:coulomb-interaction-potential}
    \frac{V(\qo{\R}_{i}, \qo{\R}_{j}, \qo{\r}_{i}, \qo{\r}_{j})}{C e^{2}} = \frac{4}{\ab{\qo{\R}_{i} - \qo{\R}_{j}}} - \frac{2}{\ab{\qo{\R}_{i} - \qo{\R}_{j} + \qo{\r}_{i}}} - \frac{2}{\ab{\qo{\R}_{i} - \qo{\R}_{j} - \qo{\r}_{j}}} + \frac{1}{\ab{\qo{\R}_{i} - \qo{\R}_{j} + \qo{\r}_{i} - \qo{\r}_{j}}},
\end{equation}
where $C = 1/4 \pi \epsilon_{0}$ denotes the Coulomb constant and $\epsilon_{0}$ the electric constant (i.e., the vacuum permittivity).
Taking this together with the expressions for the nonnegligible terms of the trapped Rydberg ion Hamiltonian, it follows that the Hamiltonian for $N$ interacting trapped Rydberg ions can be written as,
\begin{equation}
    \H = \sum_{i = 1}^{N} \H_{i} + \sum_{\mathclap{\substack{i, j = 1 \\ j < i}}}^{N} \qo{V}_{i, j},
\end{equation}
where $\H_{i} = H(\qo{\R}_{i}, \qo{\P}_{i}, \qo{\r}_{i}, \qo{\p}_{i})$ is the Hamiltonian term governing the noninteracting dynamics of the trapped ion $i$ and $\qo{V}_{i, j} = V(\qo{\R}_{i}, \qo{\R}_{j}, \qo{\r}_{i}, \qo{\r}_{j})$ is the potential term describing the interactions between the charges of ions $i$ and $j$.
Omitting the negligible center of mass contributions from the micromotion and neglecting the finite mass of the ionic core, the Hamiltonian term for ion $i$ is given by,
\begin{equation}
\begin{aligned}
    \H_{i} & = \frac{1}{2 M} \sum_{u} [\qo{P}_{i; u}^{2} + M^{2} \nu_{u}^{2} \qo{R}_{i; u}^{2}] + \frac{\qo{\p}_{i}^{2}}{2 m} + V(\ab{\qo{\r}_{i}}) - e \alpha \cos(\omega t) [\qo{r}_{i; x}^{2} - \qo{r}_{i; y}^{2}] - e \beta [3 \qo{r}_{i; z}^{2} - \qo{\r}_{i}^{2}] - e \qo{\r}_{i} \cdot \bm{E}(t) \\ & \qquad
    - 2 e \alpha \cos(\omega t) [\qo{R}_{i; x} \qo{r}_{i; x} - \qo{R}_{i; y} \qo{r}_{i; y}] - 2 e \beta [3 \qo{R}_{i; z} \qo{r}_{i; z} - \qo{\R}_{i} \cdot \qo{\r}_{i}].
\end{aligned}
\end{equation}
In order to express the Hamiltonian for the interacting trapped Rydberg ions in a form similar to that for the single trapped Rydberg ion, namely, in terms of its external, internal, and coupled motions, we must necessarily rewrite the potential term in Eq.~\eqref{eq:coulomb-interaction-potential} describing the Coulomb interaction between the charges in terms of a series by performing a multipole expansion about the center of mass relative positions $\ab{\qo{\R}_{i} - \qo{\R}_{j}}$.
For typical experimental parameters~\cite{mokhberi2020} the mean distance between the ions is expected to be significantly more than the mean distance between the ionic core and valence electron of the ions $\ex{\R_{i} - \R_{j}} \gg \ex{\r_{i}}$.
We can, therefore, accurately approximate the Coulomb interaction potential by neglecting the higher order corrections of the multipole expansion.
For simplicity, we only consider terms up to second order such that the expression for the potential term in Eq.~\eqref{eq:coulomb-interaction-potential} can be well approximated by~\cite{muller2008},
\begin{equation}
    \frac{\qo{V}_{i, j}}{C e^{2}} = \frac{1}{\ab{\qo{\R}_{ij}}} + \frac{\qo{\bm{n}}_{ij} \cdot \qo{\r}_{i}}{\ab{\qo{\R}_{ij}}^{2}} - \frac{\qo{\bm{n}}_{ij} \cdot \qo{\r}_{j}}{\ab{\qo{\R}_{ij}}^{2}} - \frac{3 [\qo{\bm{n}}_{ij} \cdot \qo{\r}_{i}]^{2} - \qo{\r}_{i}^{2}}{2 \ab{\qo{\R}_{ij}}^{3}} - \frac{3 [\qo{\bm{n}}_{ij} \cdot \qo{\r}_{i}][\qo{\bm{n}}_{ij} \cdot \qo{\r}_{j}] - \qo{\r}_{i} \cdot \qo{\r}_{j}}{\ab{\qo{\R}_{ij}}^{3}} - \frac{3 [\qo{\bm{n}}_{ij} \cdot \qo{\r}_{j}]^{2} - \qo{\r}_{j}^{2}}{2 \ab{\qo{\R}_{ij}}^{3}},
\end{equation}
where we have introduced the following notations for the center of mass relative positions,
\begin{equation}
    \qo{\bm{n}}_{ij} = \frac{\qo{\R}_{ij}}{\ab{\qo{\R}_{ij}}}, \qquad
    \qo{\R}_{ij} = \qo{\R}_{i} - \qo{\R}_{j}.
\end{equation}
The first term describes the monopole-monopole interaction, specifically, the interaction between the electric monopole moment of ion $i$ with that of ion $j$.
The second and third terms then denote the dipole-monopole and monopole-dipole interactions, namely, the interactions between the electric dipole moment of ion $i$ with the electric monopole moment of ion $j$ and, similarly, that of ion $j$ with ion $i$.
These arise due to the displacements of the orbiting valence electrons from their ionic cores which leads to the induction of electric dipole moments that interact with the electric monopole moments of the other ion.
Likewise, the fourth and sixth terms, the quadrupole-monopole and monopole-quadrupole interactions, result from the induced electric quadrupole moments of each ion interacting with the electric monopole moments of the other ion.
Note that it can be straightforwardly shown (e.g., by retaining the effective charge numbers of the ionic cores), that each of these terms vanish for the case of interacting trapped Rydberg atoms~\cite{saffman2010}.
The fifth and final term is then the well known dipole-dipole interaction.

At sufficiently low temperature $T \sim 0 \ \mathrm{K}$ and high relative trapping $\alpha \gg \beta$, the ions undergo a phase transition in which they align along the trap axis (i.e., in the $z$ direction) to form crystalline one-dimensional structures referred to as \textit{Coulomb crystals}~\cite{bollinger1994}.
In such structures, the ions vibrate about equilibrium positions determined by the interplay between the repulsive Coulomb forces between the ions and the attractive trapping forces confining the ions~\cite{james1998}.
The equilibrium positions of the ions follow from the stationary point of the potential governing the center of mass motion of the ionic cores and are calculated by solving the coupled differential equations,
\begin{equation}\label{eq:external-potential-stationary-point}
    \del_{i} \qo{V}_{\mathrm{ex}} |_{\qo{\R}_{i} = \qo{\R}_{i}^{0}} = \bm{0}, \qquad
    \qo{V}_{\mathrm{ex}} = \frac{M}{2} \sum_{i = 1}^{N} [\nu_{x}^{2} \qo{R}_{i; x}^{2} + \nu_{y}^{2} \qo{R}_{i; y}^{2} + \nu_{z}^{2} \qo{R}_{i; z}^{2}] + C e^{2} \sum_{\mathclap{\substack{i, j = 1 \\ j < i}}}^{N} \frac{1}{\ab{\qo{\R}_{ij}}},
\end{equation}
where $\qo{V}_{\mathrm{ex}}$ is the external potential and $\qo{\R}_{i}^{0}$ is the equilibrium position of ion $i$ with $\del_{i} = (\partial/\partial \qo{R}_{i; x}, \partial/\partial \qo{R}_{i; y}, \partial/\partial \qo{R}_{i; z})$.
To simplify the following calculations, we introduce the characteristic frequency scale $\nu$ and associated dimensionless trap frequencies $\gamma_{u}$ which are defined in terms of the oscillation frequencies by $\nu_{u} = \gamma_{u} \nu$ with $\gamma_{x} = \gamma_{y} = \gamma$ and $\gamma_{z} = 1$.
It then follows straightforwardly from the definitions of the oscillation frequencies in Eq.~\eqref{eq:trap-frequencies} that,
\begin{equation}
    \nu = \sqrt{\frac{4 e \beta}{M}}, \qquad
    \gamma = \sqrt{\frac{2 e^{2} \alpha^{2}}{M^{2} \omega^{2} \nu^{2}} - \frac{1}{2}},
\end{equation}
where $\alpha$, $\beta$, and $\omega$ are the radial and axial electric field gradients and radial drive frequency, respectively, and $\gamma$ is~the trap anisotropy characterizing the relative strength of the radial to axial trapping.
For $\gamma > \gamma_{*}$, where $\gamma_{*}$ is the critical value of the trap anisotropy which scales as $\gamma_{*} \sim 0.556 \smash{N^{0.915}}$~\cite{enzer2000}, the one-dimensional Coulomb crystal undergoes a second order phase transition into a two-dimensional \textit{Wigner crystal}~\cite{fishman2008}.
Here, however, we only consider the regime for which $\gamma < \gamma_{*}$ such that the equilibrium positions $\smash{\qo{\R}_{i}^{0} = (0, 0, \qo{R}_{i; z}^{0})}$.
Introducing the characteristic length scale $\zeta$ associated to the equilibrium distances between ions,
\begin{equation}\label{eq:characteristic-length}
    \zeta = \sqrt[3]{\frac{C e^{2}}{M \nu^{2}}},
\end{equation}
and the corresponding dimensionless equilibrium positions $R_{i}$, defined by $\smash{R_{i; z}^{0}} = \zeta R_{i}$, the coupled differential equations in Eq.~\eqref{eq:external-potential-stationary-point} can be succinctly recast as, 
\begin{equation}\label{eq:equilibrium-position-condition}
    R_{i} = \sum_{\mathclap{\substack{j = 1 \\ j \neq i}}}^{N} \frac{R_{ij}}{\ab{R_{ij}}^{3}} = \sum_{\mathclap{\substack{j = 1 \\ j < i}}}^{N} \frac{1}{R_{ij}^{2}} - \sum_{\mathclap{\substack{j = 1 \\ j > i}}}^{N} \frac{1}{R_{ij}^{2}},
\end{equation}
where $R_{ij} = R_{i} - R_{j}$ and with the implicit assumption that $R_{1} < R_{2} < \cdots < R_{N}$.

Following the detailed analysis in Ref.~\cite{muller2008}, we now perform a harmonic expansion of the center of mass positions of the ions about their equilibrium positions.
Retaining terms only up to second order, the Hamiltonian of the interacting trapped Rydberg ions can be conveniently expressed in terms of its external, internal, and coupled dynamics as,
\begin{equation}
    \H = \H_{\mathrm{ex}} + \H_{\mathrm{in}} + \H_{\mathrm{ex\text{-}in}}.
\end{equation}
The first term describing the center of mass motions of the ions reads,
\begin{equation}
    \H_{\mathrm{ex}} = \frac{1}{2 M} \sum_{i = 1}^{N} \sum_{u} \bigg[\qo{P}_{i; u}^{2} + M^{2} \nu^{2} \sum_{j = 1}^{N} K_{ij; u} \qo{Q}_{i; u} \qo{Q}_{j; u}\bigg],
\end{equation}
where $\qo{\Q}_{i} = \qo{\R}_{i} - \qo{\R}_{i}^{0}$ denotes the displacement of the center of mass of ion $i$ from its equilibrium position and $K_{ij; u}$ a coefficient of the Hessian matrix, which can be defined in terms of the generalized coefficients $K_{ij}$ by,
\begin{equation}
    K_{ij} =
    \begin{cases}
        \displaystyle{\gamma^{2} - K_{ij; x} = \gamma^{2} - K_{ij; y} = \frac{K_{ij; z} - 1}{2}} = \sum_{\mathclap{\substack{k = 1 \\ k \neq i}}}^{N} \frac{1}{\ab{R_{ik}}^{3}}, & \text{if $i = j$}, \\
        \displaystyle{- K_{ij; x} = - K_{ij; y} = \frac{K_{ij; z}}{2} = -\frac{1}{\ab{R_{ij}}^{3}}}, & \text{if $i \neq j$}. \\
    \end{cases}
\end{equation}
The second term dictating the relative motions of the valence electrons of the trapped Rydberg ions in the combined electric potentials of the screened ionic core, linear Paul trap, and applied fields is,
\begin{equation}
    \H_{\mathrm{in}} = \H_{\mathrm{in}}^{\mathrm{free}} + \H_{\mathrm{in}}^{\mathrm{trap}} + \H_{\mathrm{in}}^{\mathrm{field}} + \H_{\mathrm{in}}^{\mathrm{dip}}.
\end{equation}
Reminiscent of the single trapped Rydberg ion, the former term encodes the dynamics of the field-free valence electron orbiting in the modified Coulomb potential of the screened nucleus,
\begin{equation}
    \H_{\mathrm{in}}^{\mathrm{free}} = \sum_{i = 1}^{N} \bigg[\frac{\qo{\p}_{i}^{2}}{2 m} + V(\ab{\qo{\r}_{i}})\bigg].
\end{equation}
The following term describing the interactions of the valence electrons with the electric quadrupolar potential of~the linear Paul trap is modified by the monopole-quadrupole and quadrupole-monopole interactions from the multipole expansion of the Coulomb potential and reads,
\begin{equation}
    \H_{\mathrm{in}}^{\mathrm{trap}} = - e \sum_{i = 1}^{N} \big[\alpha \cos(\omega t) [\qo{r}_{i; x}^{2} - \qo{r}_{i; y}^{2}] + \beta K_{ii; z} [3 \qo{r}_{i; z}^{2} - \qo{\r}_{i}^{2}]\big].
\end{equation}
The next term dictates the interactions of the valence electrons with the electric dipolar potential of the applied field, which we assume takes the form of a polychromatic plane wave, and is written as,
\begin{equation}
    \H_{\mathrm{in}}^{\mathrm{field}} = - e \sum_{i = 1}^{N} \qo{\r}_{i} \cdot \bm{E}(t).
\end{equation}
In contrast, the final, yet familiar dipole-dipole interaction term describes the interactions between the induced dipole moments of the valence electrons of the ions and can be expressed as,
\begin{equation}
    \H_{\mathrm{in}}^{\mathrm{dip}} = \frac{M \nu^{2}}{2} \sum_{\mathclap{\substack{i, j = 1 \\ j < i}}}^{N} K_{ij; z} [3 \qo{r}_{i; z} \qo{r}_{j; z} - \qo{\r}_{i} \cdot \qo{\r}_{j}].
\end{equation}
Finally, the third term governing the coupling between the external and internal dynamics is,
\begin{equation}
    \H_{\mathrm{ex\text{-}in}} = \H_{\mathrm{ex\text{-}in}}^{\mathrm{trap}},
\end{equation}
where, similar to the internal dynamics, the term due to the electric potential of the linear Paul trap is modified by the monopole-dipole and dipole-monopole interactions of the Coulomb potential expansion and reads,
\begin{equation}
    \H_{\mathrm{ex\text{-}in}}^{\mathrm{trap}} = - 2 e \sum_{i = 1}^{N} \bigg[\alpha \cos(\omega t) [\qo{Q}_{i; x} \qo{r}_{i; x} - \qo{Q}_{i; y} \qo{r}_{i; y}] + \beta \sum_{\mathclap{j = 1}}^{N} K_{ij; z} [3 \qo{Q}_{i; z} \qo{r}_{j; z} - \qo{\Q}_{i} \cdot \qo{\r}_{j}]\bigg].
\end{equation}

From here, we proceed by introducing phonon modes via the ladder operator method, which transform the external dynamics Hamiltonian describing the center of mass motion of the ions into a diagonal form.
Accordingly, we define~the canonical coordinates, that is, the center of mass displacement and its conjugate momentum, $\qo{Q}_{i; u}$ and $\qo{P}_{i; u}$, in terms of bosonic creation and annihilation operators, $\qo{a}_{p; u}^{\dagger}$ and $\qo{a}_{p; u}$, by,
\begin{equation}\label{eq:normal-coordinate-transformation}
    \qo{Q}_{i; u} = \frac{\chi}{\sqrt{2}} \sum_{p = 1}^{N} \frac{1}{\sqrt{\gamma_{p; u}}} \Gamma_{i, p; u} [\qo{a}_{p; u}^{\dagger} + \qo{a}_{p; u}], \qquad
    \qo{P}_{i; u} = \i M \nu \frac{\chi}{\sqrt{2}} \sum_{p = 1}^{N} \sqrt{\gamma_{p; u}} \Gamma_{i, p; u} [\qo{a}_{p; u}^{\dagger} - \qo{a}_{p; u}],
\end{equation}
where we have introduced the characteristic length scale $\chi$ associated to the equilibrium oscillations of the ions,
\begin{equation}
    \chi = \sqrt{\frac{\hbar}{M \nu}}.
\end{equation}
The coefficients $\Gamma_{i, p; u}$ are elements of an orthogonal matrix that diagonalizes the Hessian matrix of coefficients $K_{ij; u}$ and as such satisfy the following defining identities in terms of the generalized coefficients $K_{ij}$ with,
\begin{equation}\label{eq:hessian-matrix-general}
    \sum_{i = 1}^{N} \Gamma_{i, p} \Gamma_{i, q} = \delta_{p, q}, \qquad
    \sum_{\mathclap{i, j = 1}}^{N} \Gamma_{i, p} K_{ij} \Gamma_{j, q} = \gamma_{p} \gamma_{q} \delta_{p, q}, \qquad
    \sum_{j = 1}^{N} K_{ij} \Gamma_{j, p} = \gamma_{p}^{2} \Gamma_{i, p},
\end{equation}
where the coefficients of the orthogonal matrices and associated dimensionless frequencies are related by,
\begin{equation}
    \Gamma_{i, p} = \Gamma_{i, p; x} = \Gamma_{i, p; y} = \Gamma_{i, p; z}, \qquad
    \gamma_{p}^{2} = \gamma^{2} - \gamma_{p; x}^{2} = \gamma^{2} - \gamma_{p; y}^{2} = \frac{\gamma_{p; z}^{2} - 1}{2}.
\end{equation}
In diagonal form, namely, in terms of the phonon modes, the Hamiltonian governing the external \textit{vibrational} dynamics can then be written as,
\begin{equation}\label{eq:external-vibrational-hamiltonian}
    \H_{\mathrm{ex}} = \hbar \nu \sum_{p = 1}^{N} \sum_{u} \gamma_{p; u} \qo{a}_{p; u}^{\dagger} \qo{a}_{p; u},
\end{equation}
where $\gamma_{p; u}$ is the dimensionless frequency of the phonon mode corresponding to the creation~and annihilation operators $\qo{a}_{p; u}$ and $\qo{a}_{p; u}^{\dagger}$, respectively. 

With the external vibrational motion resolved, we now consider the internal \textit{electronic} dynamics.
To this end, let~us introduce the energy eigenbasis of the field-free Hamiltonian, in which the relative position operators read,
\begin{equation}
    \qo{r}_{i; u}^{k} \equiv \sum_{\mu, \nu} \me{\mu}{\qo{r}_{u}^{k}}{\nu} \op{\mu}{\nu}_{i},
\end{equation}
where, for ease of notation, we have utilised the superindex quantum number $\mu \equiv (n, l, s, j, m_{j})$ for which the sum is over all states of interest.
Notice that we have dropped the explicit label $i$ of the ion on the relative position operators matrix elements since the energy eigenvalues and eigenstates of the field-free Hamiltonian are \textit{independent} of the ion, however, to avoid ambiguity we retain the label $i$ on the associated outer product to distinguish which Hilbert space the operator acts on.
In this work, we restrict our discussion to a subspace containing just three electronic states with low orbital and total angular momentum (i.e., $l \leq 1$ and $j \leq 1/2$), specifically, an energetically low-lying ground state which we denote by $\ket{0}$ and a pair of energetically high-lying Rydberg states given by $\ket{1}$ and $\ket{2}$ with~$E_{2} > E_{1} \gg E_{0}$ where $E_{\mu} \equiv E_{n, l, j}$ indicates the energy of the state $\ket{\mu} \equiv \ket{n, l, s, j, m_{j}}$ with $n$, $l$, $s$, $j$, and $m_{j}$ respectively denoting the principal, orbital angular momentum, spin angular momentum, total angular momentum,~and total magnetic~quantum numbers.
In particular, we consider the highly-excited Rydberg $\mathrm{s}$ and $\mathrm{p}$ states,
\begin{equation}
    \ket{1} \equiv \ket{n, 0, 1/2, 1/2, -1/2}, \qquad
    \ket{2} \equiv \ket{n, 1, 1/2, 1/2, 1/2},
\end{equation}
where $n \gg 1$.
The lowly-excited ground state $\ket{0}$ is then assumed to be a metastable state of the ion. Specifically, in the main text, we identify the trapped Rydberg ions as either barium ions, namely, $\smash{^{138}\mathrm{Ba}^{+}}$ ions where $\ket{0} \equiv \ket{5\mathrm{D}_{5/2}}$, or strontium $\smash{^{88}\mathrm{Sr}^{+}}$ ions where $\ket{0} \equiv \ket{4\mathrm{D}_{5/2}}$.
The ground state $\ket{0}$ is coupled to the Rydberg state $\ket{1}$ by a two-photon excitation scheme (see the review in Ref.~\cite{mokhberi2020} and references therein) via an intermediate state $\ket{7\mathrm{P}_{3/2}}$ or $\ket{6\mathrm{P}_{3/2}}$ which is easily accessible using electric dipole transitions (see Refs.~\cite{kleczewski2012, williams2013, inlek2017}). 
For simplicity, we neglect the explicit details~of the two-photon excitation scheme and instead consider the effective electric transition directly coupling these states.
The physical motivations for such a restriction are manifold.
First, these Rydberg states are experimentally the~most easily accessible from the ground state via laser excitation generated utilising established four-wave mixing techniques (see, e.g., Refs.~\cite{schmidtkaler2011, kolbe2012, bachor2016}).
Second, these states are energetically well isolated from the degenerate manifold of higher orbital angular momentum states and, moreover, are sufficiently well energetically separated from their adjacent total angular momentum states (i.e., states with identical principal and orbital angular momentum quantum numbers, but different total angular momentum quantum numbers) such that there is negligible coupling between states due to the electric quadrupolar potential of the linear Paul trap (for details see, e.g., Refs.~\cite{muller2008, schmidtkaler2011, mokhberi2020}).
Finally, states with~total orbital angular momentum quantum number $j = 1/2$ do not possess a permanent quadrupole moment and, therefore, do not experience energy level shifts due to the aforementioned electric potential of the linear Paul trap (see Ref.~\cite{higgins2017}).
Taken together, these restrictions allow us to neglect the interaction between the valence electron and linear Paul trap entirely.
Hence, the Hamiltonian describing the internal electronic dynamics, represented in the basis of the field-free Hamiltonian of the valence electron, reads,
\begin{equation}
    \H_{\mathrm{in}} = \H_{\mathrm{in}}^{\mathrm{free}} + \H_{\mathrm{in}}^{\mathrm{field}} + \H_{\mathrm{in}}^{\mathrm{dip}}.
\end{equation}

The Hamiltonian governing the field-free motion of the valence electron is, of course, diagonal and given by,
\begin{equation}
    \H_{\mathrm{in}}^{\mathrm{free}} = \sum_{i = 1}^{N} [E_{2} \op{2}{2}_{i} + E_{1} \op{1}{1}_{i} + E_{0} \op{0}{0}_{i}].
\end{equation}
For the term describing the interaction between the charge of the valence electron and electric potential of the applied field, we notice that since the Rydberg states do not possess a permanent dipole moment, the diagonal matrix elements are zero, that is, $\me{\mu}{\qo{r}_{u}}{\mu} = 0$.
To calculate the remaining nonzero electric transition dipole moments, we exploit the separability of the wavefunction to factor the states into radial and angular parts that we address independently~\cite{weber2017}.
The radial matrix elements are computed numerically using the radial wavefunctions which are obtained by solving~the associated radial Schr{\"o}dinger equation for the field-free electronic Hamiltonian while the angular matrix elements are calculated analytically using standard angular momentum algebra~\cite{louck2023}.
Explicitly calculating the latter, we find that the Hamiltonian for the electron-field interaction can be written in spherical polar coordinates as,
\begin{equation}
    \H_{\mathrm{in}}^{\mathrm{field}} = \frac{1}{3} e E_{x}(t) \sum_{i = 1}^{N} \big[\me{2}{\qo{r}}{1} [\op{2}{1}_{i} + \op{1}{2}_{i}] + \me{1}{\qo{r}}{0} [\op{1}{0}_{i} + \op{0}{1}_{i}]\big],
\end{equation}
where in order to obtain this expression we have assumed that the electric field is linearly polarized in the $x$-direction and propagating in the $z$-direction, that is, $\bm{E}(t) = (E_{x}(t), 0, 0)$ and calculated the electric transition dipole moments,
\begin{equation}
    \me{2}{\qo{\r}}{1} = - \frac{1}{3} \me{2}{\qo{r}}{1} (1, -\i, 0), \qquad
    \me{1}{\qo{\r}}{0} = - \frac{1}{3} \me{1}{\qo{r}}{0} (1, -\i, 0).
\end{equation}
Due to the angular momentum selection rules, the expressions for the dipole-dipole interaction terms can be simplified analogously.
Here, however, we can additionally exploit the fact that the electric transition dipole moment scales with the principal quantum number (see, e.g., Ref.~\cite{mokhberi2020} and references therein and the reviews in Refs.~\cite{gallagher1988, gallagher1994, gallagher2023}).
Consequently, since there is no applied electric field to compensate the relative magnitude of the electric transition dipole moments, as was the case for the electron-field interaction term, we can neglect all contributions to the dipole-dipole interaction except for those between the Rydberg states (i.e., $\ab{\me{2}{\qo{r}}{1}} \gg \ab{\me{1}{\qo{r}}{0}}$).
As such, we find the dipole-dipole interaction Hamiltonian can be well approximated by,
\begin{equation}
    \H_{\mathrm{in}}^{\mathrm{dip}} = - \frac{2}{9} M \nu^{2} \ab{\me{2}{\qo{r}}{1}}^{2} \sum_{\mathclap{\substack{i, j = 1 \\ j < i}}}^{N} K_{ij} [\op{2}{1}_{i} \op{1}{2}_{j} + \op{1}{2}_{i} \op{2}{1}_{j}].
\end{equation}
Considering now the coupled motion we find that, with the relative coordinates written in the energy eigenbasis~of~the field-free Hamiltonian of the valence electron and the center of mass coordinates written in terms of the phonon mode creation and annihilation operators, that the coupling due to the electric potential of the linear Paul trap reads,
\begin{equation}
    \H_{\mathrm{ex\text{-}in}} = \frac{\sqrt{2}}{3} e \beta \chi \me{2}{\qo{r}}{1} \sum_{i = 1}^{N} \sum_{p = 1}^{N} \frac{\Gamma_{i, p}}{\sqrt[4]{\gamma^{2} - \gamma_{\smash{p}}^{2}}} \bigg[F_{p; -}(t) [\qo{a}_{p; x}^{\dagger} + \qo{a}_{p; x}] [\op{2}{1}_{i} + \op{1}{2}_{i}] + \i F_{p; +}(t) [\qo{a}_{p; y}^{\dagger} + \qo{a}_{p; y}] [\op{2}{1}_{i} - \op{1}{2}_{i}]\bigg],
\end{equation}
where we have introduced the convenient shorthand notations for the dimensionless mode-dependent electric fields,
\begin{equation}
    F_{p; \pm}(t) = 2 \sqrt{2 \gamma^{2} + 1} \frac{\omega}{\nu} \cos(\omega t) \pm [2 \gamma_{p}^{2} + 1].
\end{equation}
With the relevant terms of the general Hamiltonian for the many-body quantum system of interacting trapped Rydberg ions defined, we are now in a position to derive the specific coupled spin-boson model Hamiltonian of interest.

\subsection{Spin-phonon coupled model Hamiltonian of a trapped Rydberg ion chain}

For typical experimental parameters (see the recent review in Ref.~\cite{mokhberi2020}), which we will employ throughout this work, the interactions between the trapped ions in Rydberg states are relatively weak compared to the energy associated to the trapping frequency $\nu$ of the external vibrational motion of the ions~\cite{li2014}.
In order to overcome this, we implement the method of Rydberg state dressing, in which a microwave (MW) frequency electric field is used to drive transitions between the Rydberg states~\cite{muller2008, li2013, li2014}.
Ultimately, this induces permanent oscillating dipole moments in the dressed Rydberg states resulting in remarkably strong and controllable dipole-dipole interactions between the ions.
Moreover, for sufficiently strong interactions, it can be shown that the external vibronic motion of the ionic core and the internal electronic motion of the valence electron approximately decouple (i.e., to zeroth order; see aforementioned references).
In this limit,~the leading order coupling arises from the first order expansion of the dipole-dipole interaction~\cite{zhang2020, magoni2023}.

The dressing of the Rydberg states is implemented by the applied electric field which, thus far, has assumed the~form of a general polychromatic plane wave.
For specificity, however, we henceforth consider a bichromatic plane wave,
\begin{equation}
    E_{x}(t) = A_{1} \cos(\omega_{1} t) + A_{2} \cos(\omega_{2} t),
\end{equation}
where $A_{\mu}$ is the electric field gradient of the plane wave with corresponding wavevector $\bm{k}_{\mu}$~and frequency $\omega_{\mu} = c \ab{\bm{k}_{\mu}}$.
To bring the Hamiltonian into a practical form, we move into the rotating frame via the unitary,
\begin{equation}
    \U_{i} = \e^{\i E_{0} t / \hbar} [\e^{\i [\omega_{2} + \omega_{1}] t} \op{2}{2}_{i} + \e^{\i \omega_{1} t} \op{1}{1}_{i} + \op{0}{0}_{i}].
\end{equation}
After performing the rotating wave approximation, whereby we neglect the rapidly oscillating time-dependent terms~in the Hamiltonian, we find that to zeroth order the external and internal dynamics decouple, that is, the coupling term is negligible.
This follows from the fact that the modulation of the radial RF driving field, despite being time-dependent, is insignificant relative to the oscillation of the applied electric fields and, as such, can be considered effectively static.
Hence, the coupled dynamics is dominated by the complex exponential phases which, for the experimental timescales under consideration, swiftly average to zero.
Consequently, in the rotating frame, the full system Hamiltonian can be readily approximated by,
\begin{equation}
    \H = \H_{\mathrm{ex}} + \H_{\mathrm{in}}, \qquad
    \H_{\mathrm{in}} = \H_{\mathrm{in}}^{\mathrm{free}} + \H_{\mathrm{in}}^{\mathrm{field}} + \H_{\mathrm{in}}^{\mathrm{dip}},
\end{equation}
where the Hamiltonian governing the external vibrational dynamics is still given by the expression in Eq.~\eqref{eq:external-vibrational-hamiltonian} whilst the Hamiltonian terms describing the internal electronic dynamics are given by (see, e.g., Ref.~\cite{muller2008}),
\begin{equation}
\begin{aligned}
    \H_{\mathrm{in}}^{\mathrm{free}} & = \hbar \sum_{i = 1}^{N} \big[[\Delta_{2} + \Delta_{1}] \op{2}{2}_{i} + \Delta_{1} \op{1}{1}_{i}\!\big], \\
    \H_{\mathrm{in}}^{\mathrm{field}} & = - \frac{\hbar}{2} \sum_{i = 1}^{N} \big[\Omega_{2}[\op{2}{1}_{i} + \op{1}{2}_{i}] - \Omega_{1}[\op{1}{0}_{i} + \op{0}{1}_{i}]\big], \\
    \H_{\mathrm{in}}^{\mathrm{dip}} & = \frac{2}{9} M \nu^{2} \ab{\me{2}{\qo{r}}{1}}^{2}  \sum_{\mathclap{\substack{i, j = 1 \\ j < i}}}^{N} \frac{1}{\ab{R_{ij}}^{3}} [\op{2}{1}_{i} \op{1}{2}_{j} + \op{1}{2}_{i} \op{2}{1}_{j}],
\end{aligned}
\end{equation}
where we have introduced the detunings $\Delta_{\mu}$ and Rabi frequencies $\Omega_{\mu} > 0$ for $\mu = 1, 2$, defined by,
\begin{equation}
    \Delta_{2} = \frac{E_{2} - E_{1}}{\hbar} - \omega_{2}, \qquad
    \Delta_{1} = \frac{E_{1} - E_{0}}{\hbar} - \omega_{1}, \qquad
    \Omega_{2} = - \frac{1}{3} \frac{e}{\hbar} \me{2}{\qo{r}}{1} A_{2}, \qquad
    \Omega_{1} = \frac{1}{3} \frac{e}{\hbar} \me{1}{\qo{r}}{0} A_{1},
\end{equation}
and recalled the definitions of the frequency $\nu$ and coefficients $K_{ij}$. We now dress the Rydberg states in the MW field of frequency $\omega_{2}$.
To manifest~this, we diagonalize the Hamiltonian for the manifold of Rydberg states, specifically,
\begin{equation}
    \H_{\mathrm{in}}^{\mathrm{Ryd}} = \hbar \sum_{i = 1}^{N} \bigg[[\Delta_{2} + \Delta_{1}] \op{2}{2}_{i} + \Delta_{1} \op{1}{1}_{i} - \frac{\Omega_{2}}{2} [\op{2}{1}_{i} + \op{1}{2}_{i}]\bigg],
\end{equation}
the resulting eigenvalues $\hbar \Delta_{\pm}$ and corresponding orthonormalized eigenvectors $\ket{\pm}$ of which are given by,
\begin{equation}
    \Delta_{\pm} = \Delta_{1} + \frac{\Delta_{2} \pm \sqrt{\Delta_{\smash{2}}^{2} + \Omega_{\smash{2}}^{2}}}{2}, \qquad
    \ket{\pm} = \frac{N_{\pm}}{\sqrt{2}} \ket{2} \pm \frac{N_{\mp}}{\sqrt{2}} \ket{1}, \qquad
    N_{\pm} = \pm \sqrt{1 \pm \frac{\Delta_{\smash{2}}}{\sqrt{\Delta_{\smash{2}}^{2} + \Omega_{\smash{2}}^{2}}}}.
\end{equation}
For the experimental parameters used here, the energy splitting between the MW dressed Rydberg states is sufficiently large such that we can neglect the off-resonant coupling of the laser field of frequency $\omega_{1}$ to the higher energy dressed Rydberg state $\ket{+}$.
Under this assumption, we obtain an effective two-level system consisting of the low-lying ground state $\ket{0}$ and high-lying dressed Rydberg state $\ket{-}$ with corresponding effective Hamiltonian,
\begin{equation}
    \H = \hbar \nu \sum_{p = 1}^{N} \gamma_{p; u} \qo{a}_{p; u}^{\dagger} \qo{a}_{p; u} + \hbar \sum_{i = 1}^{N} \bigg[\Delta_{-} \qo{n}_{i} + \frac{\Omega_{-}}{2} \qo{\sigma}_{i}^{x}\bigg] + \frac{1}{9} N_{+}^{2} N_{-}^{2} M \nu^{2} \ab{\me{2}{\qo{r}}{1}}^{2} \sum_{\mathclap{\substack{i, j = 1 \\ j < i}}}^{N} \frac{1}{\ab{R_{ij}}^{3}} \qo{n}_{i} \qo{n}_{j},
\end{equation}
where we have additionally introduced the dressed Rabi frequencies $\Omega_{\pm}$, Rydberg state occupation operators $\qo{n}_{i}$, and Pauli operators $\qo{\sigma}_{i}^{u}$ given by,
\begin{equation}
    \Omega_{\pm} = \mp \frac{N_{\mp}}{\sqrt{2}} \Omega_{1}, \qquad
    \qo{n}_{i} = \op{-}{-}_{i}, \qquad
    \qo{\sigma}_{i}^{x} = \op{-}{0}_{i} + \op{0}{-}_{i}.
\end{equation}
In general, the (dressed) Rydberg states exhibit strongly enhanced electric polarizabilities that scale with the principal quantum number as $n^{7}$ (see, e.g., the review in Ref.~\cite{mokhberi2020} and references therein) that derive from the coupled electronic and vibrational motion of the valence electron and ionic core due to the presence of the electric potential of the linear Paul trap~\cite{li2013, li2014}.
Consequently, the \textit{undressed} Rydberg states $\ket{1}$ and $\ket{2}$ experience modified state dependent trap frequencies $\nu_{u} \to \nu_{\mu; u}$ and, therefore, exhibit both state and mode dependent phonon frequencies $\nu_{p; u} \to \nu_{p; \mu; u}$ where $\nu \equiv \gamma_{u} \nu$ and $\nu_{p; u} \equiv \gamma_{p; u} \nu$.
For an appropriate choice of the electric MW field parameters $\Delta_{2}$ and $\Omega_{2}$, the polarizability of the \textit{dressed} Rydberg states $\ket{\pm}$ can, however, be tailored such that the energy shifts due to the polarizability of the undressed Rydberg states vanishes (for details, see the discussion in Ref.~\cite{li2014}), thus eliminating the state dependence.
To achieve this requires a particular choice of Rydberg states and electric field parameters that is not necessary for~the present work, since the coupling has already been eliminated via the rotating wave approximation, but will necessarily prove essential to take into consideration in future investigations.
For our purposes, it is sufficient to consider the~case in which the dipole-dipole interaction strengths are \textit{maximised} which occurs when the detuning of the MW frequency electric field vanishes.
In this limit, $\Delta_{2} = 0$, we find that $N_{\pm} = \pm 1$ and so,
\begin{equation}
    \Delta_{\pm} = \Delta_{1} \pm \frac{\Omega_{2}}{2}, \qquad
    \Omega_{\pm} = \frac{\Omega_{1}}{\sqrt{2}}, \qquad
    \ket{\pm} = \pm \frac{1}{\sqrt{2}} [\ket{2} \mp \ket{1}].
\end{equation}

In the rotating frame (and under the rotating wave approximation), the external vibrational and internal electronic dynamics become decoupled and, as such, the effective Hamiltonian is block diagonal.
Here, the leading order coupling arises from the further expansion of the dipole-dipole interaction potential (see Refs.~\cite{zhang2020, magoni2023}).
Specifically, expanding the center of mass positions about their equilibrium positions to first order in their respective displacements we obtain,
\begin{equation}
    \frac{1}{\ab{R_{ij}}^{3}} \approx \frac{1}{\ab{R_{ij}}^{3}} - \frac{3}{\zeta \ab{R_{ij}}^{4}} \qo{Q}_{ij; z}, \qquad
    \zeta = \sqrt[3]{\frac{C e^{2}}{M \nu^{2}}},
\end{equation}
where we have remarked that $j < i$, so $R_{ij} > 0$, and, hence, $R_{ij} = \ab{R_{ij}}$ with $\qo{Q}_{ij; z} = \qo{Q}_{i; z} - \qo{Q}_{j; z}$ the displacements of the ions about their equilibrium positions and $\zeta$ the characteristic length scale associated to the equilibrium distances between the ions (see Eq.~\eqref{eq:characteristic-length}).
Represented in terms of the phonon mode creation and annihilation operators, $\qo{a}_{p; z}^{\dagger}$ and $\qo{a}_{p; z}$, this first order expansion then reads,
\begin{equation}
    \frac{1}{\ab{R_{ij}}^{3}} \approx \frac{1}{\ab{R_{ij}}^{3}} - \frac{3 \chi}{\sqrt{2} \zeta \ab{R_{ij}}^{4}} \sum_{p = 1}^{N} \frac{\Gamma_{ij, p}}{\sqrt[4]{2 \gamma_{\smash{p}}^{2} + 1}} [\qo{a}_{p; z}^{\dagger} + \qo{a}_{p; z}], \qquad
    \chi = \sqrt{\frac{\hbar}{M \nu}},
\end{equation}
where $\chi$ is the characteristic length scale associated to the equilibrium oscillations of the ions and $\Gamma_{ij, p} = \Gamma_{i, p} - \Gamma_{j, p}$ are the eigenvector components associated to the phonon mode creation and annihilation operators $\qo{a}_{p; z}^{\dagger}$ and $\qo{a}_{p; z}$.
It then follows that the spin-phonon coupled model Hamiltonian can be written compactly as, 
\begin{equation}
    \H = \hbar \nu \sum_{p = 1}^{N} \gamma_{p; z} \qo{a}_{p; z}^{\dagger}\qo{a}_{p; z} + \hbar \Delta_{-} \sum_{i = 1}^{N} \qo{n}_{i} + \frac{\hbar \Omega_{-}}{2} \sum_{i = 1}^{N} \qo{\sigma}_{i}^{x} + \sum_{\mathclap{\substack{i, j = 1 \\ j < i}}}^{N} V_{ij} \qo{n}_{i} \qo{n}_{j} +  \sum_{\mathclap{\substack{i, j, p = 1 \\ j < i}}}^{N} \kappa_{ij, p} [\qo{a}_{p; z}^{\dagger} + \qo{a}_{p; z}] \qo{n}_{i} \qo{n}_{j},
\end{equation}
where the interaction strengths~$V_{ij}$ and coupling strengths~$\kappa_{ij, p}$ are defined by,
\begin{equation}
    V_{ij} \equiv M \nu^{2} \frac{\ab{\me{2}{\qo{r}}{1}}^{2}}{9 \ab{R_{ij}}^{3}}, \qquad
    \kappa_{ij, p} \equiv - M \nu^{2} \sqrt{\frac{\hbar}{2 M \nu}} \sqrt[3]{\frac{M \nu^{\smash{2}}}{C e^{2}}} \frac{\ab{\me{2}{\qo{r}}{1}}^{2}}{3 \ab{R_{ij}}^{4}} \frac{\Gamma_{ij, p}}{\sqrt[4]{2 \gamma_{\smash{p}}^{2} + 1}},
\end{equation}
where for brevity we note that the relative strength scales with the characteristic length scales as $\kappa_{ij, p}/V_{ij} \sim \chi/\zeta \ll 1$.
We can, however, even further simplify the notation by remarking that we are only interested in the centermost pair~of ions of the trapped Rydberg ion chain.
Explicitly, if all but the centermost pair of ions are in the ground state of the system, i.e., $\ket{0}$, then their dynamics decouple from the chain, since the interaction and coupling terms vanish.~Hence,
if we introduce the dimensionless equilibrium distance between the centermost pair of ions, which corresponds to the minimum distance between any pair of ions~\cite{james1998}, denoted by,
\begin{equation}
    R \equiv \min_{i, j}(|R_{ij}|),
\end{equation}
then we can, similarly, define the associated mode components $\Gamma_{p} \equiv \Gamma_{ij, p}$,~interaction strength $V \equiv V_{ij}$, and coupling strengths $\kappa_{p} \equiv \kappa_{ij, p}$.
Since we only consider the axial modes (i.e., the radial modes are decoupled), we can additionally introduce the following standard notations for the mode frequencies $\omega_{p} \equiv \nu_{p; z} = \gamma_{p; z} \nu$ and operators $\qo{a}_{p} \equiv \qo{a}_{p; z}$.
From here, we also introduce the following simplified expressions for the detuning $\Delta$ and Rabi frequency $\Omega$,
\begin{equation}
    \Delta \equiv \Delta_{-} = \Delta_{1} - \frac{\Omega_{2}}{2}, \qquad
    \Omega \equiv \frac{\Omega_{-}}{2} = \frac{\Omega_{1}}{\sqrt{8}}.
\end{equation}
Taking this all together, we attain the desired form for the spin-phonon coupled model Hamiltonian for the centermost pair of trapped Rydberg ions, equivalent to Eq.~(8) of the main text, which reads,
\begin{equation}\label{eq:model-hamiltonian}
    \H = \hbar \sum_{p = 1}^{N} \omega_{p} \qo{a}_{p}^{\dagger}\qo{a}_{p} + \hbar \Delta [\qo{n}_{1} + \qo{n}_{2}] + \hbar \Omega [\qo{\sigma}_{1}^{x} + \qo{\sigma}_{2}^{x}] + V \qo{n}_{1} \qo{n}_{2} +  \sum_{p = 1}^{N} \kappa_{p} [\qo{a}_{p}^{\dagger} + \qo{a}_{p}] \qo{n}_{1} \qo{n}_{2},
\end{equation}
where, for convenience, we have labelled the centermost pair of ions with $i = 1, 2$.
Finally, introducing more standard notation for the electric transition dipole moment $d$ and equilibrium distance between the centermost pair~of ions $R_{0}$,
\begin{equation}
    d \equiv - \frac{e \me{2}{\qo{r}}{1}}{3}, \qquad
    R_{0} \equiv \zeta R,
\end{equation}
we get that the interaction strength $V$ and coupling strengths $\kappa_{p}$ take the more conventional form (cf. Eq.~(2)),
\begin{equation}
    V \equiv \frac{1}{4 \pi \epsilon_{0}} \frac{d^{2}}{R_{0}^{3}}, \qquad
    \kappa_{p} \equiv - \frac{1}{4 \pi \epsilon_{0}} \frac{3 d^{2}}{R_{0}^{4}} \sqrt{\frac{\hbar}{2 M \omega_{\smash{p}}}} \Gamma_{p}.
\end{equation}

\subsection{Quantum Rabi model Hamiltonian of a trapped Rydberg ion chain}

In order to probe the spectrum of the spin-phonon coupled system, we consider the facilitation regime, wherein the interaction energy $V$ is cancelled by the laser detuning $\Delta$ (i.e., we set $\hbar \Delta = - V$).
For the case of $N = 2$ ions, the laser only couples the unexcited ground state $\ket{\downarrow\downarrow}$ to the singly excited symmetric state $\ket{\mathcal{S}} = [\ket{\uparrow\downarrow} + \ket{\downarrow\uparrow}] / 2$ and the doubly excited Rydberg state $\ket{\uparrow\uparrow}$.
To manifest this, we transform the Hamiltonian in Eq.~\eqref{eq:model-hamiltonian}, which for $N = 2$ is represented in matrix form in the effective spin basis $\{\ket{\uparrow\uparrow}, \ket{\uparrow\downarrow}, \ket{\downarrow\uparrow}, \ket{\downarrow\downarrow}\}$ as (n.b., we set $\hbar = 1$ henceforth),
\begin{equation}
    \H = 
    \begin{bmatrix}
        -V & \Omega & \Omega & 0 \\
        \Omega & -V & 0 & \Omega \\
        \Omega & 0 & -V & \Omega \\
        0 & \Omega & \Omega & 0 \\
    \end{bmatrix}
    +
    \begin{bmatrix}
        \kappa_{2} & 0 & 0 & 0 \\
        0 & 0 & 0 & 0 \\
        0 & 0 & 0 & 0 \\
        0 & 0 & 0 & 0 \\
    \end{bmatrix}
    \! [\qo{a}_{2}^{\dagger} + \qo{a}_{2}] + \omega_{2} \qo{a}_{2}^{\dagger} \qo{a}_{2},
\end{equation}
where we have neglected the decoupled center of mass motion.
Employing the aforementioned unitary transformation which maps the basis states of the electronic spin degrees of freedom as $\{\ket{\uparrow\uparrow}, \ket{\uparrow\downarrow}, \ket{\downarrow\uparrow}, \ket{\downarrow\downarrow}\} \mapsto \{\ket{\uparrow\uparrow}, \ket{\mathcal{S}}, \ket{\mathcal{A}}, \ket{\downarrow\downarrow}\}$ where $\ket{\mathcal{S}} = [\ket{\uparrow\downarrow} + \ket{\downarrow\uparrow}] / 2$ and $\ket{\mathcal{A}} = [\ket{\uparrow\downarrow} - \ket{\downarrow\uparrow}] / 2$ represent the singly excited symmetric and antisymmetric states, we obtain the transformed spin-phonon coupled model Hamiltonian which reads,
\begin{equation}
    \H = 
    \begin{bmatrix}
        -V & \sqrt{2} \Omega & 0 & 0 \\
        \sqrt{2} \Omega & -V & 0 & \sqrt{2} \Omega \\
        0 & 0 & -V & 0 \\
        0 & \sqrt{2} \Omega & 0 & 0 \\
    \end{bmatrix}
    + 
    \begin{bmatrix}
        \kappa_{2} & 0 & 0 & 0 \\
        0 & 0 & 0 & 0 \\
        0 & 0 & 0 & 0 \\
        0 & 0 & 0 & 0 \\
    \end{bmatrix}
    \! [\qo{a}_{2}^{\dagger} + \qo{a}_{2}] + \omega_{2} \qo{a}_{2}^{\dagger} \qo{a}_{2}.
\end{equation}
Neglecting the evidently decoupled singly excited antisymmetric state $\ket{\mathcal{A}}$, and noting that we will be considering the strong interaction regime for which $V \gg \Omega$, such that we can additionally omit the far off resonant unexcited ground state $\ket{\downarrow\downarrow}$, the approximate spin-phonon coupled model Hamiltonian for two ions can be rewritten as (cf. Eq.~(4)),
\begin{equation}
    \H = 
    \begin{bmatrix}
        -V & \sqrt{2} \Omega \\
        \sqrt{2} \Omega & -V \\
    \end{bmatrix}
    + 
    \begin{bmatrix}
        \kappa_{2} & 0 \\
        0 & 0 \\
    \end{bmatrix}
    \! [\qo{a}_{2}^{\dagger} + \qo{a}_{2}] + \omega_{2} \qo{a}_{2}^{\dagger} \qo{a}_{2}.
\end{equation}
For the experimental parameters considered in the main text, the interaction strength $V$ is significantly stronger than the coupling strength $\kappa_{2}$ and, therefore, we can treat the coupling term as a \textit{perturbation}.
For $\kappa_{2} = 0$, the dynamics~of the effective spins and phonon mode clearly decouple and the corresponding model Hamiltonian becomes diagonal.~In
particular, we can write the approximate model Hamiltonian $\H$ as,
\begin{equation}
    \H = \H_{0} + \delta \H,
\end{equation}
where $\H_{0}$ denotes the unperturbed Hamiltonian comprised of the decoupled spin and phonon terms, such that $\H = \H_{0}$ for $\kappa_{2} = 0$, and $\delta \H$ the perturbation that couples the dynamics.
Applying the unitary transformation that diagonalizes the unperturbed Hamiltonian $\H_{0}$ which transforms the electronic basis states of the effective spin degrees of freedom as $\{\ket{\uparrow\uparrow}, \ket{\mathcal{S}}\} \mapsto \{\ket{+}, \ket{-}\}$ where $\ket{\pm} = [\ket{\uparrow\uparrow} \pm \ket{\mathcal{S}}] / \sqrt{2}$, the Hamiltonian terms read (cf. Eq.~(5)),
\begin{equation}
    \H_{0} =
    \begin{bmatrix}
        E_{+} & 0 \\
        0 & E_{-} \\
    \end{bmatrix}
    + \omega_{2} \qo{a}_{2}^{\dagger} \qo{a}_{2}, \qquad
    \delta \H = \frac{\kappa_{2}}{2}
    \begin{bmatrix}
        1 & 1 \\
        1 & 1 \\
    \end{bmatrix}
    \! [\qo{a}_{2}^{\dagger} + \qo{a}_{2}],
\end{equation}
where the electronic energy eigenvalues $E_{\pm} = -V \pm \sqrt{2} \Omega$.
Given that the spin and phonon dynamics are decoupled~in the unperturbed Hamiltonian $\H_{0}$, its energy eigenvalues $E_{\pm, N}$ and associated eigenstates $\ket{\pm, N}$ follow as,
\begin{equation}
    E_{\pm, N} = E_{\pm} + N \omega_{2}, \qquad
    \ket{\pm, N} = \ket{\pm} \otimes \ket{N},
\end{equation}
where $\ket{N}$ denotes an eigenstate of the relative phonon mode number operator with vibrational energy eigenvalue~$N$.
Note that the number operator eigenvalue $N$ should not be confused with the number of trapped ions $N = 2$, which is fixed here.
It follows that a resonance occurs whenever any of the energy eigenvalues $E_{\pm, N}$ become degenerate, in particular, when $E_{-, N} = E_{+, M}$ for $N > M$ which occurs at the resonant Rabi frequency $\Omega = \Omega_{N\! M}^{\mathrm{res}} \approx [N - M] \omega_{2} / 2 \sqrt{2}$.
Note that this value for the resonant Rabi frequency is only an approximation, since we are neglecting all higher order corrections.
For the resonance shown in Fig.~2 of the main text, where $N = M + 1 = 1$, the approximate Hamiltonian near the resonance (i.e., at $\Omega = \Omega_{1 0}^{\mathrm{res}} \equiv \Omega_{\mathrm{res}} \approx \omega_{2} / 2 \sqrt{2}$) to first order in perturbation theory is given by,
\begin{equation}
    \H \approx \H_{0} +
    \begin{bmatrix}
        \me{-, 1}{\delta \H}{-, 1} & \me{-, 1}{\delta \H}{+, 0} \\
        \me{+, 0}{\delta \H}{-, 1} & \me{+, 0}{\delta \H}{+, 0} \\
    \end{bmatrix}
    =
    \begin{bmatrix}
        -V - \omega_{2} / 2 & \kappa_{2} / 2 \\
        \kappa_{2} / 2 & -V + \omega_{2} / 2 \\
    \end{bmatrix}.
\end{equation}
The approximate eigenvalues $E_{\pm}^{\mathrm{res}}$ and associated eigenstates $\ket{E_{\pm}^{\mathrm{res}}}$ which are given in Eq.~(6) of the main text are,
\begin{equation}
    E_{\pm}^{\mathrm{res}} \approx -V + \frac{\omega_{2}}{2} \pm \frac{\kappa_{2}}{2}, \qquad
    \ket{E_{\pm}^{\mathrm{res}}} \approx \frac{1}{\sqrt{2}} [\ket{-, 1} \pm \ket{+, 0}] = \frac{1}{2} [\ket{\uparrow\uparrow} \otimes [\ket{1} \pm \ket{0}] - \ket{\mathcal{S}} \otimes [\ket{1} \mp \ket{0}]],
\end{equation}
which clearly demonstrates the hybridization of the electronic spin and vibrational phonon degrees of freedom.
Notice that the energy splitting at the resonance is, to first order, given by the spin-phonon coupling strength $\kappa_{2}$.

\section{Theoretical quantities and experimental parameters}\label{sec:phonons}

In order to numerically simulate and investigate the spectral signatures of the spin-phonon coupling, we necessarily need explicit values for the theoretical quantities and experimental parameters in the model Hamiltonian in Eq.~\eqref{eq:model-hamiltonian}.
In Tab.~\ref{tab:parameters}, we present values for the dimensionless equilibrium separation between the centermost ions $R = R_{0} / \zeta$ and the dimensionless mode frequencies $\gamma_{p; z} = \omega_{p} / \nu$ and associated coupling coefficients $\Gamma_{p}$ for a chain of $N = 2, 4, \ldots, 20$ ions.
The remaining parameters, specifically, the isotopic masses of the ions $M$ and electric transition dipole moments between the Rydberg states $\me{2}{\qo{r}}{1} = -3 d / e$ used in the main text are,
\begin{equation}
\begin{aligned}
    ^{88}\mathrm{Sr}^{+}\mathrm{:} & \qquad
    M = 87.9\,u, & \quad
    \me{2}{\qo{r}}{1} = -1434\,a, \\
    ^{138}\mathrm{Ba}^{+}\mathrm{:} & \qquad
    M = 137.9\,u, & \quad
    \me{2}{\qo{r}}{1} = -1320\,a,
\end{aligned}
\end{equation}
where $a \approx 5.292 \times 10^{-11}\,\mathrm{m}$ is the Bohr radius and $u \approx 1.661 \times 10^{-27}\,\mathrm{kg}$ the unified atomic mass unit.

\begin{table}[ht]
    \footnotesize
    \setlength{\tabcolsep}{6pt}
    \begin{center}
    \begin{tabular}{D{.}{.}{0} D{.}{.}{3} D{.}{.}{0} D{.}{.}{3} D{.}{.}{3}}
        \hline\hline
        \multicolumn{1}{c}{$N$} & \multicolumn{1}{c}{$R_{0} / \zeta$} & \multicolumn{1}{c}{$p$} & \multicolumn{1}{c}{$\omega_{p} / \nu$} & \multicolumn{1}{c}{$\Gamma_{p}$} \\
        \hline
         2 & 1.260 &  2 &  1.732 & 1.414 \\ [4pt]
         4 & 0.909 &  2 &  1.732 & 0.426 \\
           &       &  4 &  3.051 & 1.348 \\ [4pt]
         6 & 0.740 &  2 &  1.732 & 0.224 \\
           &       &  4 &  3.058 & 0.556 \\
           &       &  6 &  4.274 & 1.281 \\ [4pt]
         8 & 0.636 &  2 &  1.732 & 0.143 \\
           &       &  4 &  3.063 & 0.329 \\
           &       &  6 &  4.286 & 0.608 \\
           &       &  8 &  5.443 & 1.225 \\ [4pt]
        10 & 0.564 &  2 &  1.732 & 0.101 \\
           &       &  4 &  3.067 & 0.225 \\
           &       &  6 &  4.296 & 0.388 \\
           &       &  8 &  5.458 & 0.631 \\
           &       & 10 &  6.576 & 1.179 \\ [4pt]
        12 & 0.511 &  2 &  1.732 & 0.076 \\
           &       &  4 &  3.070 & 0.166 \\
           &       &  6 &  4.303 & 0.277 \\
           &       &  8 &  5.471 & 0.423 \\
           &       & 10 &  6.593 & 0.640 \\
           &       & 12 &  7.682 & 1.141 \\ [4pt]
        14 & 0.469 &  2 &  1.732 & 0.060 \\
           &       &  4 &  3.073 & 0.129 \\
           &       &  6 &  4.310 & 0.211 \\
           &       &  8 &  5.482 & 0.312 \\
           &       & 10 &  6.608 & 0.445 \\
           &       & 12 &  7.701 & 0.643 \\
           &       & 14 &  8.767 & 1.107 \\
        \hline\hline
    \end{tabular}
    \begin{tabular}{D{.}{.}{0} D{.}{.}{3} D{.}{.}{0} D{.}{.}{3} D{.}{.}{3}}
        \hline\hline
        \multicolumn{1}{c}{$N$} & \multicolumn{1}{c}{$R_{0} / \zeta$} & \multicolumn{1}{c}{$p$} & \multicolumn{1}{c}{$\omega_{p} / \nu$} & \multicolumn{1}{c}{$\Gamma_{p}$} \\
        \hline
        16 & 0.436 &  2 &  1.732 & 0.049 \\
           &       &  4 &  3.075 & 0.104 \\
           &       &  6 &  4.316 & 0.167 \\
           &       &  8 &  5.492 & 0.243 \\
           &       & 10 &  6.622 & 0.337 \\
           &       & 12 &  7.718 & 0.459 \\
           &       & 14 &  8.787 & 0.642 \\
           &       & 16 &  9.834 & 1.078 \\ [4pt]
        18 & 0.408 &  2 &  1.732 & 0.041 \\
           &       &  4 &  3.077 & 0.086 \\
           &       &  6 &  4.321 & 0.137 \\
           &       &  8 &  5.500 & 0.196 \\
           &       & 10 &  6.634 & 0.267 \\
           &       & 12 &  7.733 & 0.354 \\
           &       & 14 &  8.805 & 0.468 \\
           &       & 16 &  9.856 & 0.639 \\
           &       & 18 & 10.887 & 1.053 \\ [4pt]
        20 & 0.384 &  2 &  1.732 & 0.035 \\
           &       &  4 &  3.079 & 0.073 \\
           &       &  6 &  4.326 & 0.115 \\
           &       &  8 &  5.508 & 0.163 \\
           &       & 10 &  6.644 & 0.219 \\
           &       & 12 &  7.747 & 0.285 \\
           &       & 14 &  8.822 & 0.367 \\
           &       & 16 &  9.875 & 0.474 \\
           &       & 18 & 10.910 & 0.635 \\
           &       & 20 & 11.928 & 1.030 \\ % [25.5pt]
           & & & $\vdots$ & \\ [7.8pt]
        \hline\hline
    \end{tabular}
    \end{center}
    \caption{Numerical values for the dimensionless equilibrium separation between the centermost ions $R_{0} / \zeta$ and the dimensionless mode frequencies $\omega_{p} / \nu$ and associated mode coupling coefficients $\Gamma_{p}$ for a chain of $N = 2, 4, \ldots, 20$ trapped Rydberg ions.}
    \label{tab:parameters}
\end{table}

% the following figure is included to ensure that Overleaf compiles a pdfLaTeX file (not a LaTeX2e file)

\begin{figure}
    \centering
    \includegraphics{fig5.png}
\end{figure}
\vskip -20pt

% -------------------------------------------
% [-] Bibliography
% -------------------------------------------

\bibliography{supplemental.bib}